\documentclass[]{aa}

\usepackage{txfonts}

\usepackage{color}

\begin{document}

\title{The vertical structure of the boundary layer around compact objects}

\author{Marius Hertfelder\thanks{\email{marius.hertfelder@gmail.com}}\inst{1}
        \and
                Wilhelm Kley\inst{1}
        }

   \institute{Institut für Astronomie und Astrophysik, Abt.~Computational Physics, Auf der Morgenstelle 10, 72076 Tübingen, Germany
    }

\titlerunning{The vertical structure of the BL}
\authorrunning{M.~Hertfelder and W.~Kley}

\date{Received 22 March 2017 / Accepted 11 May 2017}

\abstract
        {%
                Mass transfer due to Roche lobe overflow leads to the formation of an accretion disk around a weakly magnetized white dwarf (WD)
                in cataclysmic variables. At the inner edge of the disk, the gas comes upon the surface of the WD and has to get rid of its
                excess kinetic energy in order to settle down on the more slowly rotating outer   stellar layers. This region is known as the
                boundary layer (BL).
        }
        {%
        In this work we investigate the vertical structure of the BL, which is still poorly understood. We shall provide details of the basic structure of the two-dimensional (2D) BL and how it depends on parameters such as stellar mass and rotation rate, as well as
        the mass-accretion rate. We further investigate the destination of the disk material and compare our results with previous
        one-dimensional (1D) simulations.
        }
        {%
        We solve the 2D equations of radiation hydrodynamics in a spherical ($r$-$\vartheta$) geometry using a parallel grid-based code that employs a Riemann solver. The radiation energy is considered in the two-temperature
        approach with a radiative flux given by the flux-limited diffusion approximation.
        }
        {
                The BL around a non-rotating WD is characterized by a steep drop in angular velocity over a width of only $1\%$ of the
                stellar radius, a heavy depletion of mass, and a high temperature ($\sim 500\,000$ K) as a consequence of the strong shear.
                Variations in $\Omega_\ast,M_\ast, $ and $ \dot{M}$ influence the extent of the changes of the variables in the BL but not the general
                structure. Depending on $\Omega_\ast$, the disk material travels up to the poles or is halted at a certain latitude. The extent
                of mixing with the stellar material also depends on $\Omega_\ast$. We find that the 1D approximation matches the 2D data
                well, apart from an underestimated temperature.
        }


\maketitle

\section{Introduction}\label{sec:introduction}

Accretion of matter onto a non-magnetized star is accompanied by the formation of a boundary layer (BL) where the
Keplerian rotation rate of the disk smoothly connects to the stellar rotation rate, that is, in general, much smaller.
In the BL, up to $\sim 50\%$ of the total accretion luminosity is liberated in a spatially very confined region.
In this publication we focus on the BL around a white dwarf (WD) that resides in a cataclysmic variable (CV) system. 
The BL around a WD typically has a radial
extent of $\lesssim 1\%$ of the stellar radius \citep{1974MNRAS.168..603L,2013A&A...560A..56H}. The extent of the energy released in the BL
depends on the rotation rate of the WD. When the gas enters the BL from the outer parts of the disk, it gradually gets
rid of one half of the accretion energy during its inward drift through the disk. The other half of the energy gained by
falling into the gravitational potential of the star is stored in terms of kinetic energy. The difference of the kinetic energy
at this point and on the surface of the star of each gas particle is liberated in the BL and radiated away in the form of
soft and hard X-ray and UV emission \citep[e.g.,][]{1981MNRAS.196....1C,1981ApJ...245..609C,1984MNRAS.206..879C}. Therefore, the
slower the star spins, the higher the energy output of the BL. Apart from the stellar mass and radius, another parameter
that heavily influences the radiation characteristics of the BL is the mass-accretion rate $\dot{M}$, which can vary over a
wide range for CVs. If it is relatively low, $\dot{M}\leq 10^{-10} M_{\sun}/\text{yr}$ \citep{1987MNRAS.227...23W}, the BL is optically thin
and very high temperatures of the order $10^8\,\text{K}$ are reached, resulting in soft and hard X-ray emission (e.g., 
\citealt{2004RMxAC..20..244M,2003MNRAS.346.1231P,2005ApJ...626..396P}, 
\citealt{1984Natur.308..519K,1987Ap&SS.130..303S,1993Natur.362..820N,1999MNRAS.308..979P}). Higher mass-accretion rates lead
to optically thick BLs with lower temperatures of about $10^5\,\text{K}$ \citep{1977MNRAS.178..195P, 1986SvAL...12..117S, 1995ApJ...442..337P}
in which case the radiation is mostly thermalized (see e.g., \citealt{1980MNRAS.190...87C,2004RMxAC..20..174M}).

From a numerical point of view, the BL problem has initially been approached by timescale estimates and
stationary calculations \citep{1974MNRAS.168..603L,1977MNRAS.178..195P,1977AcA....27..235T,1981AcA....31..267T,1979MNRAS.187..777P,1983A&A...126..146R}.
With increasing computational power, evolutionary simulations became viable and made the investigation of the temporal evolution
and non-stationary phenomena possible \citep{1986MNRAS.221..279R,1987A&A...172..124K,1989A&A...208...98K,1989A&A...222..141K,1991A&A...247...95K}.
\citet{1995MNRAS.275.1093G} performed one-dimensional simulations, where axial symmetry is assumed and the equations are integrated over
the vertical ($z$) direction. This procedure is known as the thin-disk approach and provides a good first approximation for the
BL problem. In a previous paper series we performed (one-dimensional) 1D simulations including vertical cooling, radiation transport, and the
treatment of the radiation energy, and calculated detailed emergent spectra from these models \citep{2013A&A...560A..56H,2014a&a...571a..55s}.

In this paper we consider the problem in two-dimensional (2D) spherical coordinates. As in the 1D case, we still assume axisymmetry in order to
get rid of the $\varphi$-coordinate. We conduct full radiation hydrodynamics simulations, where the radiation energy is treated in the
two-temperature approximation \citep{2010A&A...511A..81K,2011A&A...529A..35C}, which means that an additional evolutionary equation
for the radiation energy is considered. Since the BLs we study here are mostly optically thick, we employ the flux-limited diffusion
approximation \citep{1981ApJ...248..321L,1984JQSRT..31..149L} for the radiative flux. We use realistic boundary conditions with a
vertically spread mass inflow at the outer edge of the simulation domain (disk) and a sophisticated initial model, which is constructed
from different 1D codes (see Sec.~\ref{sec:solution-strategy}). Our studies are hence related to the work by
\citet{1989A&A...208...98K,1989A&A...222..141K,1991A&A...247...95K} but can span much longer dynamic timescales and feature a
much higher resolution due to the parallel code and the available computing power. Multidimensional simulations of the BL have also
been done by \citet{2005ApJ...635L..69F,2006NewAR..50..509F}, although only for the adiabatic case, or \citet{2009ApJ...702.1536B},
who used a simplified energy dissipation function without radiation transport. \citet{1996ApJ...461..933K,1999ApJ...518..833K}  also did
full radiation hydrodynamics simulations for protostars and studied the evolution of FU Orionis outbursts in protostellar disks,
and \citet{2008MNRAS.386.1038B} studied the BL around neutron stars in low-mass X-ray binaries. Magnetic fields have been included in 2D by
\citet{2003ApJ...589..397K} and in three dimensions (3D) by \citet{2002MNRAS.330..895A}. In both cases, however, only low resolution and short dynamical
timescales have been presented. \citet{2012MNRAS.421...63R}, for example, performed 3D magnetohydrodynamical simulations and
also considered magnetospheric accretion but did not include radiation transport.

Simulations in the disk plane have been performed by \citet{2012ApJ...760...22B,2013ApJ...770...67B,2013ApJ...770...68B,2015A&A...579A..54H,2016ApJ...817...62P,2017ApJ...835..238B}
in order to investigate the transport of mass and angular momentum (AM) in the BL. Since it has been shown that due to the increase
of the angular velocity $\Omega$ with radius, the magnetorotational instability \citep[MRI,][]{Velikhov_1959,1960PNAS...46..253C}
is ineffective in the BL \citep{1995MNRAS.277..157G,1996MNRAS.281L..21A,2012ApJ...751...48P},
alternative concepts are currently being looked into. A promising candidate is the AM transport by acoustic waves  excited by
the sonic instability \citep{1988MNRAS.231..795G,2012ApJ...752..115B}.

A different approach to describe the interface between the star and the disk was pursued by \citet{1999AstL...25..269I,2010AstL...36..848I}
who introduced the concept of the spreading layer (SL). The SL is a 1D, vertical approximation for the deceleration
of disk material on the stellar surface, which is connected to the disk by a 2D transition region. The main difference with the radial
BL model is, that the rotating gas is spread on the star where it is then decelerated by turbulent interaction with colder stellar layers beneath it.
As a consequence, the SL radiation emerges from two rings on the northern and southern hemisphere of the star. This concept was
originally developed for neutron stars and later adapted to WDs in CVs \citep{2004ApJ...610..977P,2004ApJ...616L.155P}.
\citep{2006MNRAS.369.2036S} extended the model for the effects of general relativity and different chemical compositions of the
accreted matter.

In this publication, we want to study the vertical structure and the long-term evolution of the BL around a WD. A high resolution
is employed in order to monitor the polar spreading and mixing of the disk material on the stellar surface. We decipher whether
or not the results show indications of a SL and compare the midplane profiles with 1D simulations as presented in \citet{2013A&A...560A..56H}.

The paper is organized as follows. In Sect.~\ref{sec:physics} we present the basic physical model and list the equations used
for production and postprocessing. Section~\ref{sec:numerics} is dedicated to the numerical background of our work. We
describe the code, boundary and initial conditions and how we constructed the models, which are shown and detailed in Sect.~\ref{sec:results}.
We finish with Sect.~\ref{sec:conclusion}, where we discuss the results and come to a conclusion.

\section{Physics}\label{sec:physics}

In this section, we present the physical foundations for the simulations we have performed and that are described
later in this publication. In order to investigate the vertical structure of the BL we use the Navier-Stokes
equations in spherical coordinates ($r$, $\vartheta$, $\varphi$) and assume symmetry in the azimuthal direction.
We therefore drop all derivatives with respect to the azimuthal coordinate $\varphi$ but do hold on to the momentum
equation in the corresponding direction.

\subsection{Vertical structure equations}

The conservation of mass is represented by the continuity equation, which is given by
\begin{equation}
        \frac{\partial\rho}{\partial t} + \nabla\cdot(\rho\vec{u})
        = 0,
        \label{eq:continuity}
\end{equation}
where $\rho$ and $\vec{u}=(u_r, u_\vartheta)$ are the gas density and velocity, respectively.
The equations for the momentum conservation are conveniently expressed in terms of the physically conserved quantities $s$, $g,$ and $h$.
We therefore introduce the radial momentum density $s=\rho u_r$, the polar momentum density $g=\rho r u_\vartheta$
and the angular momentum density $h=\rho r u_\varphi \sin\vartheta = \Omega r^2\sin^2\vartheta$. With regards to these variables, the
momentum equations read
\begin{equation}
        \frac{\partial s}{\partial t} + \nabla\cdot(s\vec{u}) = \rho\left(\frac{u_\vartheta^2}{r} + \frac{u_\varphi^2}{r}\right)
        -\frac{\partial p}{\partial r} + \nabla\cdot\vec{\sigma}_r + \frac{1}{r}\sigma_{rr} - \rho\frac{\partial \Psi}{\partial r}\label{eq:momentum-r}
\end{equation}
for the radial direction,
\begin{equation}
        \frac{\partial g}{\partial t} + \nabla\cdot(g\vec{u}) = \rho \cot\vartheta u_\varphi^2 - \frac{\partial p}{\partial\vartheta}
        + \nabla\cdot(r\vec{\sigma}_\vartheta) - \cot\vartheta\sigma_{\varphi\varphi}\label{eq:momentum-theta}
\end{equation}
for the polar direction and
\begin{equation}
        \frac{\partial h}{\partial t} + \nabla\cdot(h\vec{u}) = \nabla\cdot(r\sin\vartheta\vec{\sigma}_\varphi)\label{eq:momentum-phi}
\end{equation}
for the azimuthal direction. In the above equations $p$ denotes the gas pressure $p = \frac{\rho R_\text{G} T}{\mu}$, where
$R_\text{G}=\frac{k_\text{B}}{m_\text{H}}$ with Boltzmann's constant $k_\text{B}$ and the mass of hydrogen $m_\text{H}$, $T$
is the gas temperature and $\mu$ is the mean molecular weight. $\Psi=-\frac{GM_\ast}{r}$ is the gravitational potential and
$G$ and $M_\ast$ are the gravitational constant and the mass of the star, respectively.

The viscous stress tensor is denoted by $\sigma$ and can be displayed in a coordinate-system-independent way by means
of the covariant formulation
\begin{equation}
        \sigma_{ij} = \eta \left(u_{i\,;\,j} + u_{j\,;\,i} - \frac{2}{3} u^k_{\ ;\,k} g_{ij} \right),\label{eq:stress-tensor}
\end{equation}
where $u_{m\,;\,n}$ and $u^m_{\ ;\,n}$ are the co- and contravariant derivative of the $m$-th component of the velocity
with respect to $n$, while $g_{ij}$ is the metrical tensor of the coordinate system and $\eta$ is the dynamic viscosity. We note that Eq.~(\ref{eq:stress-tensor}) gives
the covariant components of the stress tensor that have to be translated to physical components by employing the
metric coefficients. The same applies for the co- and contravariant formulation of the velocity. For details of this procedure
as well as an insight into tensor calculus we refer the interested reader to classical textbooks such as \citet{2005mmp..book.....A}
or \citet{1984oup..book.....M}. The latter book also contains a summary of the individual physical components of $\sigma$ in
spherical coordinates. The vector representations of the stress tensor in Eqs.~(\ref{eq:momentum-r}-\ref{eq:momentum-phi})
are given by $\vec{\sigma}_r=(\sigma_{rr},\sigma_{r\vartheta},\sigma_{r\varphi})$ ($\vec{\sigma}_\vartheta$ and $\vec{\sigma}_\varphi$
similarly) and the divergence is calculated accordingly.

The equation for the conservation of energy is given by
\begin{equation}
        \frac{\partial e}{\partial t} + \nabla\cdot(e\vec{u}) = -p\nabla\cdot\vec{u} + \Phi -
        \kappa_\text{P}\rho c \left(a_\text{R}T^4 - E\right),\label{eq:energy}
\end{equation}
where $e = \rho\varepsilon = \rho c_V T$ is the internal energy of the gas ($\varepsilon$ and $c_V$ are the specific internal
energy and heat capacity, respectively), and $\Phi$ is the viscous dissipation function which relates to the stress tensor
via $\Phi=(\sigma\nabla)\vec{u} = \frac{1}{2\eta}\text{Tr}\,(\sigma^2)$ \citep[see e.g.,][]{1984oup..book.....M}. The last term
in brackets on the right-hand side of equation Eq.~(\ref{eq:energy}) accounts for the heating and cooling through the energy
exchange between the gas and the radiation field due to emission and absorption processes, and $E$ denotes the radiation energy
density, which is also evolved in time in our model. The according equation reads:
\begin{equation}
        \frac{\partial E}{\partial t} + \nabla\cdot\vec{F} = \kappa_\text{P}\rho c\left(a_\text{R}T^4 - E\right)\label{eq:radiation}
.\end{equation}
Here, $\vec{F}$ is the radiative flux and $\kappa_\text{P}, c,$ and $a_\text{R}$ are the Planck mean opacity, the speed of light,
and the radiation constant. The approach that incorporates two coupled equations for the gas energy (\ref{eq:energy}) and
the radiation energy density (\ref{eq:radiation}), respectively, is called the two-temperature approximation \citep{2010A&A...511A..81K,
2011A&A...529A..35C}.

In order to close the set of equations, we need an assumption for the radiative flux $\vec{F}$. Here we use the flux-limited
diffusion approximation \citep[][FLD]{1981ApJ...248..321L,1984JQSRT..31..149L} where the radiative flux is given by
\begin{equation}
        \vec{F} = -\frac{c\lambda}{\kappa_\text{R}\rho}\nabla E,\label{eq:radiative-flux}
\end{equation}
with the Rosseland mean opacity $\kappa_\text{R}$ and the flux limiter $\lambda$. The flux limiter mediates between optically
thick regions, where radiation and matter are in thermal equilibrium, and optically thin regions, where photons can travel much
farther before they interact with matter. In our simulations we utilize the formulation of the flux limiter given
by \citet{1978JQSRT..20..541M} where $\lambda$ is calculated by
\begin{equation}
        \lambda(R) = \begin{cases}
                                        \frac{2}{3+\sqrt{9+12 R^2}}     &       0 \leq R \leq \frac{3}{2} \\
                                        \frac{1}{1+R+\sqrt{1+2R}}       &       \frac{3}{2} \leq R \leq \infty
                                \end{cases},
\end{equation}
with
\begin{equation}
        R = \frac{\left|\nabla E\right|}{\kappa_\text{R}\rho E}.
\end{equation}

The Rosseland mean opacity is further determined following Kramer's law,
\begin{equation}
        \kappa = \kappa_0 \left(\frac{\rho}{\text{g cm}^{-3}}\right)\left(\frac{T}{\text{K}}\right)^{-3.5},
\end{equation}
where $\kappa_0 = 5\times10^{24}\,\text{cm}^2\,\text{g}^{-1}$, which we also use for the Planck mean opacity \citep[see also][]{2013A&A...549A.124B}. As a lower threshold, we consider
a constant opacity $\kappa_\text{Thomson} = 0.4\,\text{cm}^2\,\text{g}^{-1}$ which accounts for
free-electron scattering processes at very high temperatures (Thomson scattering).

\subsection{Viscosity}\label{sec:viscosity}
As has been discussed in Sect.~\ref{sec:introduction}, the driving mechanism for angular momentum transport in the BL is still
a matter of ongoing research. We do not yet have a suitable prescription to implement the wave-mediated AM transport in the
equations of hydrodynamics without directly simulating it. This, however, is not feasible for the kind of simulations presented
in this publication, considering the huge demand of computational resources for highly resolved, three dimensional radiation
hydrodynamics simulations \citep[see e.g.,][]{2015A&A...579A..54H}.
We therefore employ the classic $\alpha$-prescription by \citet{1973A&A....24..337S}, which considers turbulent stresses in the
accretion disk responsible for the observed AM transport and parametrizes them according to
\begin{equation}
        \nu = \alpha c_\text{s} H,\label{eq:viscosity-nu}
\end{equation}
where $\nu = \eta/\rho$ is the kinematic viscosity and $c_\text{s}=\sqrt{\gamma\frac{p}{\rho}}$ is the sound speed. $H$ is a length scale that
stands for the maximum eddy size of the turbulence. We define it according to \citet{1986MNRAS.220..593P}:
\begin{equation}
        \frac{1}{H^2} = \frac{1}{H_\text{d}^2} + \frac{1}{H_r^2}\label{eq:viscosity-H}
.\end{equation}
By employing Eq.~(\ref{eq:viscosity-H}) for the length scale, we account for the fact that in the BL the radial pressure scale height
$H_r$ becomes smaller than the vertical one. The relations for the scale heights read
\begin{equation}
        H_\text{d} = \frac{c_\text{s}}{\Omega_\text{K}}\quad\text{and}\quad H_r = \frac{p}{\left|\text{d}p/\text{d}r\right|},
\end{equation}
where $\Omega_\text{K} = \sqrt{GM_\ast/r^3}$ is the Keplerian angular velocity. Inside the star, which is included, to a lesser extent, in our simulations, we apply a constant small viscosity $\nu_\text{const} = 10^{12} \text{cm}^2/\text{s}$.
In the low-density region above the disk, no viscous heating occurs.


\subsection{One-dimensional disks and stellar structure equations}

In Sect.~\ref{sec:numerics} we describe in detail the workflow we used in order to create the vertical models of
the BL. Apart from the actual 2D simulations, the initial conditions also involve 1D disk profiles as well as a simple
representation of the star. The underlying equations shall be presented here in short.

The 1D disk profiles are created using the thin disk approximation where the Navier-Stokes equations are vertically integrated
and azimuthal symmetry is assumed. The 3D equations are thus reduced to depend only on the time $t$ and the
cylindrical radius $R = r\cdot\sin\vartheta$. The mass density $\rho$ is replaced by the surface density $\Sigma$ during
the vertical integration. Since a Gaussian profile is assumed for the vertical density stratification of the disk, we can
recalculate the midplane mass density from the results of the 1D simulations and approximate the vertical density and
temperature structure. The equations of the slim disk approach and details about the code we used to obtain the 1D profiles
can be found in \citet{2013A&A...560A..56H}.

The outermost layers of the star are obtained from the radial hydrostatic equilibrium and flux conservation:
\begin{gather}
        \frac{\text{d}p}{\text{d}r} = - \frac{\rho G M_\ast}{r^2} \label{eq:star-hydro}\\
        r^2 F = R_\ast^2 \sigma_\text{SB} T_\ast^4 = \text{const.}\label{eq:star-flux}
\end{gather}
Here, $F = F_\text{rad} + F_\text{conv}$ is the stellar flux, which consists of a radiative and a convective part, $T_\ast$
is the effective temperature of the star (i.e., $T$ at the optical depth $\tau = 1$) and $\sigma_\text{SB}$ is the 
Stefan-Boltzmann constant. All other quantities are identical to the ones described earlier. We neglect
the convective flux in Eq.~(\ref{eq:star-flux}) since we found that it is sufficient to only use the radiative flux to
obtain a robust starting model for the star that can be plugged into the full 2D simulation. The two stellar structure
equations are coupled ordinary differential equations that can be solved for $\rho(r)$ and $T(r)$. The radius of the star
$R_\ast$ is determined from the mass-radius relation for white dwarfs by \citet{1972ApJ...175..417N}.

\section{Numerics}\label{sec:numerics}

\subsection{General remarks}

We mainly used the code \texttt{PLUTO} \citep{2007ApJS..170..228M} for the simulations presented in this publication. In \texttt{PLUTO},
the conservation laws Eqs.~(\ref{eq:continuity},\ref{eq:momentum-r}\,-\,\ref{eq:momentum-phi},\ref{eq:energy}) are discretized on a static grid using the finite
volume (FV) formalism where volume averages evolve in time. A high-resolution shock-capturing (HRSC) scheme is applied, with the
algorithm being based on a reconstruct--solve--average (RSA) strategy. This means, that the variables are interpolated to the cell
interfaces where subsequently a Riemann problem is solved for the two discontinuous states. Finally, the system is evolved in time
using the fluxes computed by the Riemann solver. In order to approximate the variables at the cell interfaces, we used the
piecewise parabolic method (PPM, 5\textsuperscript{th} order), which is known to handle curvilinear coordinates and non-uniform grid spacing more correctly
than other implemented methods\footnote{%
See \url{plutocode.ph.unito.it/files/userguide.pdf}
}. The code retains a global 2\textsuperscript{nd}-order accuracy as fluxes
are computed at the interface midpoints. \texttt{PLUTO} offers a variety of slope limiters for the reconstruction step that
differ in diffusivity. We have found that for our problem it is adept to start with a more diffusive limiter (minmod-limiter) and
switch to the least diffusive monotonized central difference limiter (MC-limiter) once the simulation converges to the equilibrium.
The Riemann problem is solved using the approximative HLLC solver, which is a Harten, Lax, Van Leer method that also considers
the contact discontinuity. Finally, a dimensionally unsplit third-order TVD Runge-Kutta scheme with a variable time step based on
the Courant-Friedrichs-Lewy condition \citep{Courant28} is employed for the temporal integration. The code has been tested and used
extensively for a variety of applications\footnote{See \url{plutocode.ph.unito.it/Publications.html}.}.

Since we also considered the evolution of the radiation energy density (Eq.~\ref{eq:radiation}), we made use of the additional module 
by \citet{2013A&A...559A..80K}. Here, an additional step is performed after the non-radiative part has been accomplished.
In this step, the radiation energy is solved together with the corresponding
absorption-emission term in the gas energy equation. This results in a system of two coupled differential equations \citep[see also][]{2011A&A...529A..35C}. Since
radiation processes typically happen on much shorter timescales than hydrodynamical processes, an explicit treatment of these
equations would confine the time step severely and render the method impractical. Therefore, an implicit scheme is applied to
handle the radiative part. The resulting system of equations, which is partially linearized following \citet{2011A&A...529A..35C}, is solved using the matrix solver from the PETSc
library. For our problem, the improved stabilized version of the biconjugate gradient squared method (KSPIBCGS) in connection
with a block jacobian or \texttt{hypre} preconditioner has turned out to be a fast and reliable combination. A variety of tests have
been performed in order to confirm the correctness of the implementation and applicability of the method \citep{2013A&A...559A..80K}.
Due to the fact that the radiation module as described above can only handle 3D problem setups, we have modified it
to support one and two dimensions, as well, and have redone the test cases.

\subsection{Initial conditions}\label{sec:solution-strategy}

During the preparations and the first runs we noticed that \texttt{PLUTO} is quite susceptible to crashes due to negative
density or pressure. In the BL problem, large density contrasts appear between the star-disk system and the virtually empty space
above the disk, where high radial infall velocities are reached. However, the closer the initial model\footnote{In general, the boundary
conditions play a more important role within the Navier-Stokes equations than the initial conditions.} matches the equilibrium
state, the more stable the simulations are. Thus, we developed a strategy where we build the full problem step by step from
smaller subproblems and perpetually construct initial models from the results of the preceding step. We outline this approach below:

We begin by constructing the very outer layers of the star. For that purpose the stellar structure Eqs.~(\ref{eq:star-hydro},
\ref{eq:star-flux}) are solved for $\rho(r)$ and $T(r)$ using a fourth-order Runge-Kutta integrator. The integration starts at $r=1$
and is performed outwards. The initial density $\rho(r=1)$ is varied until we obtain $T=T_\ast$ at optical depth $\tau=1$. Then, the
integration starts at $r=1$ again and is performed inwards. The simulation domain of the star spans $[0.99,1.01]$ and a large number
of grid cells is needed due to the large gradients.

The following step entails calculating a radial 1D profile for the disk; we use our self-developed code, which solves the partial differential
equations on a fixed Eulerian, staggered grid using a semi-implicit-explicit time stepping scheme with operator splitting
\citep[see also][]{1992ApJS...80..753S}. In contrast to the simulations performed in \citet{2013A&A...560A..56H}
we do not simulate the BL, but only the disk for this step. Therefore we shift the inner boundary to $r=1.1$ and require the azimuthal
velocity to be Keplerian while we impose zero-gradient boundary conditions for all other variables. The outer boundary resides
at $r=5$ and the BCs remain unchanged. After the 1D disk has reached its equilibrium state we switch to the \texttt{PLUTO} code,
load in the 1D profiles and extrapolate them by using a Gaussian vertical stratification. The disk simulation is now 2D
with a domain that spans from $\vartheta=0$ to $\pi/2$. We run the \texttt{PLUTO} code until the 2D disk 
has reached a stationary state.

Finally, we join together the ingredients mentioned above. The domain now ranges from $0.99$ to $5$ in radius with 1118 logarithmically spaced grid cells and from $0$ to
$\pi/2$ with 1000 grid cells in polar angle. The number of cells has been chosen such that each cell is approximately quadratic. 
The stellar structure model and the 2D disk are loaded in and the small region $[1,1.1]$ in radius that has
no yet been simulated is simply extrapolated from the disk. In this region, the BL starts to build immediately after starting the
full 2D simulation.

We have chosen the aforementioned procedure since it minimizes the time effort and maximizes the chances for a successful run without
crashes. Another obvious way to pursue would be to simulate 1D BL profiles and attach these to the star and extend them to 2D.
This approach is, however, much more time consuming and the initial model is less stable.

\subsection{Boundary conditions and density floor}\label{sec:bcs}

At the outer boundary at $r = 5,$ mass is continuously entering the domain, representing the mass flux $\dot{M}$ through the disk. Numerically
this is accomplished by manipulating the flux into the outermost active cells that is actually calculated by the Riemann solver. We spread the mass
influx over a vertical extent assuming a Gaussian profile whose width is determined through the disk height that comes out of the 1D disk
simulation. The influx is realized such that the mass-accretion rate $\dot{M}$ is a simple runtime input parameter. The inflowing
matter further receives an azimuthal momentum so that $u_\varphi$ matches the Keplerian velocity at the outer boundary. 
The radiation energy density at the outer boundary is fixed at a value corresponding to $T = 100$ K. In the disk, we impose a no-flux
condition since the radial transport is very small.
The other quantities are implemented with zero-gradient BCs. Above the disk, the outer radial boundary is closed to prevent additional mass from
falling in.

The inner boundary at $r=0.99$ is a rigid wall where the radial and polar velocities equal zero. The azimuthal velocity is implemented as an
input parameter so we can easily change the rotation rate of the star $\Omega_\ast$.
For the radiation energy density, the radiative flux of the star at the inner domain edge is used as a boundary condition.
All other variables are again implemented using zero-gradient or Neuman-type boundary conditions, which means that the normal derivative at the boundary vanishes. 
At the polar axis ($\vartheta = 0$) and in the equatorial plane ($\vartheta=\pi/2$) we apply symmetry conditions (vanishing gradients) as boundary conditions.

We apply a lower threshold for the density of $5\times10^{-14} \text{g}/\text{cm}^3$ , which is approximately ten orders of magnitude smaller than the disk midplane
density. It is effective in the empty area above the disk and prevents negative densities due to infalling material. The
density floor, however, also causes a flat density structure in this region. Along with the small radiation energy density above the
disk, the computation of the flux-limiter becomes inaccurate and flux barriers develop that prevent the cooling of the disk. Since
this process is unphysical, we reset the flux-limiter to $1/3$ in areas of low $\rho$ and $E$ in order to maintain a radiative flux
towards the domain boundary. Since the ambient region is dynamically unimportant due to its low density, this intervention does not
affect the validity of the model.

\subsection{Model parameters}\label{sec:sims}

In this publication, we are interested in the structure of the BL around a weakly magnetized white dwarf, a situation that is frequently found in cataclysmic
variable systems. White dwarfs in CVs typically have masses in the range of one solar mass and effective temperatures of $T_\ast\sim50\,000\,\text{K}$
\citep[e.g.,][for SS Cygni]{2010ApJ...716L.157S}. We leave the effective temperature fixed at $50\,000\,\text{K}$. The radial scale
height of the star grows with $T_\ast$ and it is easier to simulate a star with a larger scale height since a lower resolution may be chosen.
The mass of the WD, on the other hand, is one parameter that we have varied during our research. Analogous to \citet{2013A&A...560A..56H}, we
considered three masses, $M_\ast = 0.8M_\sun$, $1.0M_\sun$  and $1.2M_\sun$. Alongside the mass, the radius of the WD is the crucial parameter
that determines the strength of the gravitational pull. We used the mass-radius relation from \citet{1972ApJ...175..417N} in order to
calculate $R_\ast$, which, for instance, yields $R_\ast \approx 5.6\times10^{8}\,\text{cm}$ for $M_\ast=1.0 M_\sun$. An equatorial radius increase
for rapidly rotating WDs is automatically taken into account in the simulations. The WD rotation rate $\Omega_\ast$ itself is an important
parameter that we vary from $0.0\Omega_\text{K}$ (non-rotating) up to $0.9\Omega_\text{K}$ for fast rotating stars. $\Omega_\ast$ 
determines the amount of kinetic energy that the gas loses before meeting the star and therefore the total luminosity of the BL. 
The last parameter that we modify is the mass-accretion rate $\dot{M}$ , which can comprise several orders of magnitude in CVs. We consider
mass-accretion rates of $\dot{M} = (10^{-8}-10^{-10})\,M_\sun/\text{yr}$. Furthermore, we took $\alpha = 0.01$ for the viscosity parameter and
$\gamma = 5/3$ (monoatomic ideal gas) for the adiabatic index. The gas is assumed to consist of completely ionized hydrogen, so we adopt
$\mu = 0.5$ for the mean molecular weight.

\section{Results}\label{sec:results}

\subsection{The basic structure of the 2D BL}\label{sec:basic-structure}

We begin the presentation of the results by discussing the general structure and basic properties of the 2D BL. For this
purpose we adopt the parameter set of $M_\ast=0.8 M_\sun, \dot{M}=10^{-8} M_\sun/\text{yr},\Omega_\ast=0.0\Omega_\text{K}$ as our reference
model which serves as the starting point for our analysis. The measured mass-accretion rate of the reference model is nearly
constant throughout the domain. The maximum deviation from the imposed value is below $10\%$. This also applies to the simulations mentioned
later in the text. Typically the models reach a state where the general flow structure remains steady after some tens of orbits. It does,
however, take a few hundred orbits for the temperature to settle down to the equilibrium value and for the mass-accretion rate to
exactly match the imposed value. All models presented in this paper are virtually in a steady state.

\begin{figure*}[t]
\begin{center}
\includegraphics[width=\textwidth]{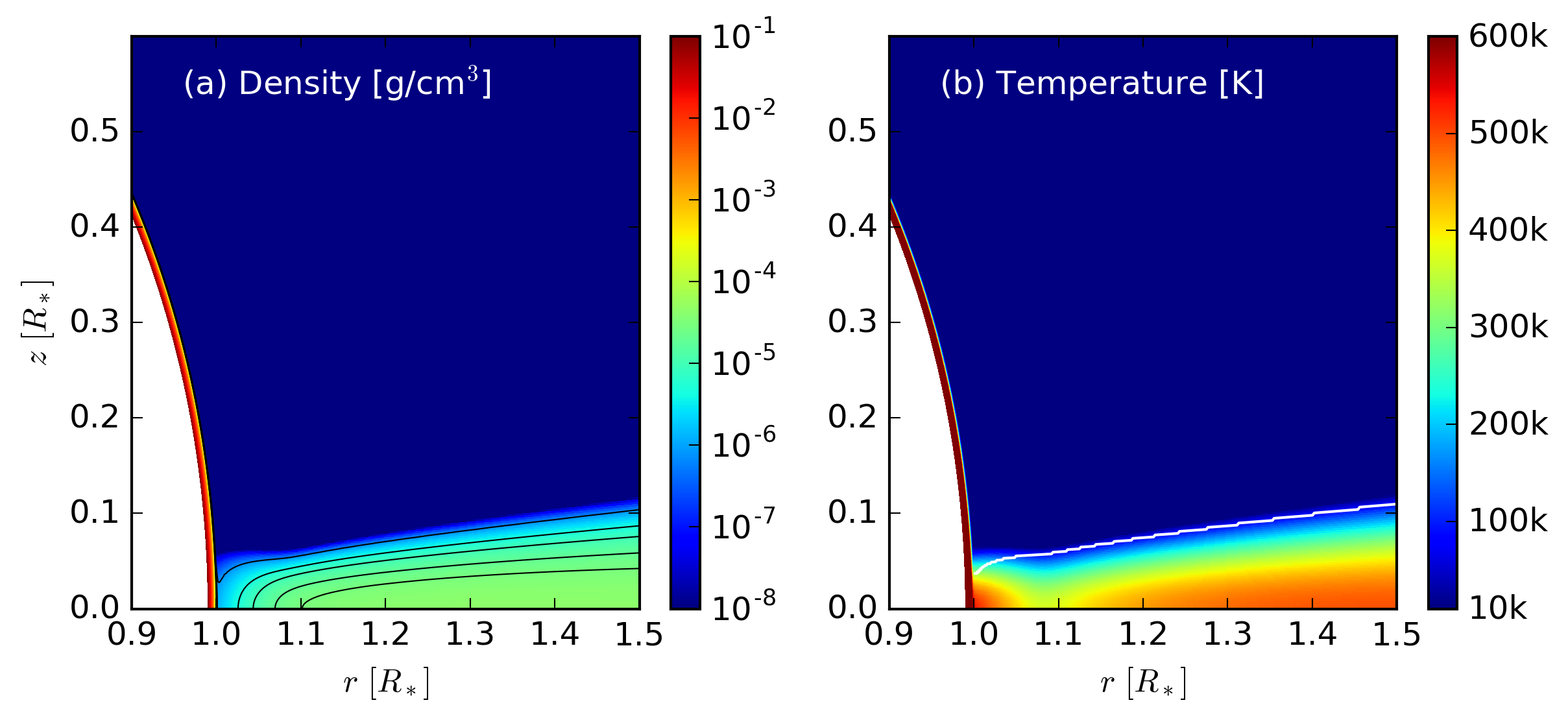}
\caption{\label{fig:rho-T-ref}%
Density (a) and temperature (b) for the reference model as a function of radius $r$ and vertical coordinate $z$ in cgs units at orbit 688. Contour lines for the levels
$(0.8, 5, 10, 20, 30)\times 10^{-6} \text{g}/\text{cm}^3$ are overlaid in the plot of the density (a). The white line in panel (b) denotes
the points where the optical depth is unity, $\tau=1$.
}
\end{center}
\end{figure*}

The thermal structure of the BL after 688 orbits is illustrated in Fig.~\ref{fig:rho-T-ref} with the mass density $\rho$ on the left hand side (a) and the
temperature $T$ on the right hand side (b). Additional contour lines are added to panel (a) in order to clarify the structure of the disk.
We first discuss the density structure of the reference model. A heavy depletion of mass is observed directly in front of the stellar
equator and the density decreases by approximately two orders of magnitude compared to the disk. This region, which connects the disk with
the stellar surface, is called the BL. Due to the decreasing density and increasing infall velocity, it resembles a bottleneck where
matter has to go through before it can come to rest on the star. At the surface of the star, the density rises rapidly with decreasing
distance from the center and is several orders of magnitude larger than in the disk, which begins on the other side of the BL. The density
in the disk is highest in the equatorial plane and decreases like a Gauss function in the vertical direction until it reaches
the density floor. This area is represented in 
the plot by the dark blue color-coding. The gas from the disk accumulates in an equatorial shell around the star, which is barely visible 
since it is very thin. This is an indication for efficient radiative cooling of the gas that is heated up in the BL. The thin layer of disk
material floats on the stellar surface and slowly spreads towards the poles. After almost 700 orbits it reaches a latitude of $\sim35^\circ$
and continues to crawl towards the pole.
One orbit is given by the time needed for a point on the stellar surface to make a full rotation, $T=2\pi/\Omega_\text{K}(R_\ast)\approx 11.5\,\text{s}$.

\begin{figure}[t]
\begin{center}
\includegraphics[width=0.5\textwidth]{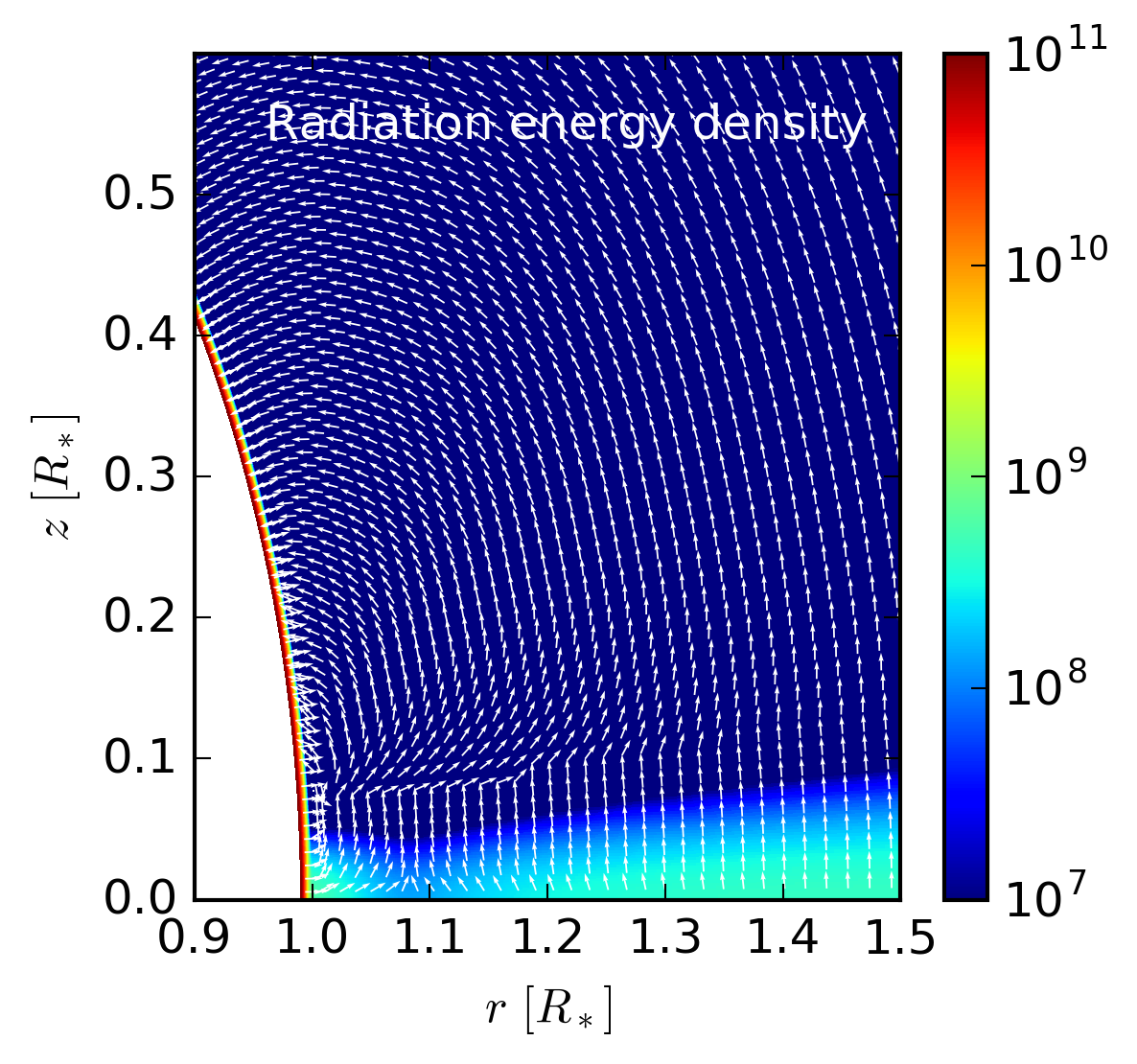}
\caption{\label{fig:Erad-ref}%
Radiation energy density $E$ of the reference model in $\text{g}/(\text{cm}\,\text{s}^2)$ as a function of radius and height for the inner part of the accretion disk
at orbit 688.
The arrows denote the direction of the radiative flux as specified in Eq.~(\ref{eq:radiative-flux}).
}
\end{center}
\end{figure}

The temperature distribution of the gas is visualized in Fig.~\ref{fig:rho-T-ref} (b). The main feature is the peak in the equatorial plane
at $r\approx 1$ with a temperature of almost $550\,000$ K. It is the consequence of the strong shearing in the BL due to the abrupt
drop of the angular velocity. The shearing component of the stress tensor is the dominant part for the dissipation, and the heat production
is proportional to the square of the shear. At the maximum of the angular velocity at $r=1.0065$, the shear vanishes since the gradient is zero,
$\partial\Omega/\partial r = 0$ (\emph{zero-torque point}). Thus the heat production declines for greater radii $r\gtrsim 1$ before it rises
again in the accretion disk. This minimum of $T$ is visible at $r\approx 1.1$ in the equatorial plane in Fig.~\ref{fig:rho-T-ref} (b). The
location of the minimum does, however, not coincide with the zero-torque point because of the transport of energy mainly through
radiation. Moreover, the cooling rate strongly depends on the density, which changes drastically in this region and thus also influences
the temperature distribution. Another important point is that due to the high radial velocities in this area, the components $\sigma_{rr}$
and $\sigma_{\varphi\varphi}$ also become important concerning the heat production. It is therefore a complex interplay of several factors
that arranges the temperature here. In the disk, beyond $r=1.1,$ the midplane temperature increases again because of the ever-present shear of a differentially
rotating accretion disk. It peaks at $r=1.51$ and $T=495\,857$ K and decreases again for larger radii with a slope approximately proportional
to $r^{-1/2}$ due to the influence of the BL. For larger radii $r\gg 1$, $T$ approaches the $r^{-3/4}$ variation of the standard disk solution.
In the vertical direction, the disk can cool efficiently and hence the temperature drops rapidly to a few thousand K. The
temperature of the low-density material above the disk decreases with radius.

The radiation energy density $E$ shows the same features as the gas temperature since the medium is mainly optically thick in the disk and BL. The energy 
dissipation in the BL also leads to a drastic increase in radiation energy density. In addition to $E$, Fig.~\ref{fig:Erad-ref} visualizes
the radiative flux $\vec{F}$ which is computed according to Eq.~(\ref{eq:radiative-flux}). At around $r\approx 1.1$ there is
also a minimum in $E$ that lies between the BL and the disk in the equatorial plane. Due to the gradient of the radiation energy density inwards
and outwards from this minimum, there is considerable radiation transport towards the point $r\approx 1.1$. The energy produced in the BL
is therefore transported outwards by radiation. Also some part of the energy produced in the disk is transported inwards.
Thus there is a significant redistribution of energy both from the BL and from the disk to a wider region of about 10 percent of the
stellar radius in radial extent. This region is called the \emph{thermal BL} \citep{1995MNRAS.272...71R,1995ApJ...442..337P}.
In the disk, the cooling from the surface is accounted for by the radiation transport in the vertical direction, which is indicated by the nearly
vertical arrows pointing upwards in Fig.~\ref{fig:Erad-ref}. Directly above the BL, the radiative flux is very strong and pushes the disk
surface flux outwards. At higher stellar latitudes, Fig.~\ref{fig:Erad-ref} shows that the flux is directed towards the star.
This unexpected behavior is probably a consequence of the small density in the corona and the numerical problems for the radiation transport
associated with the density floor. However, this region does not influence the structure or the thermodynamics of the BL.

\begin{figure*}[t]
\begin{center}
\includegraphics[width=\textwidth]{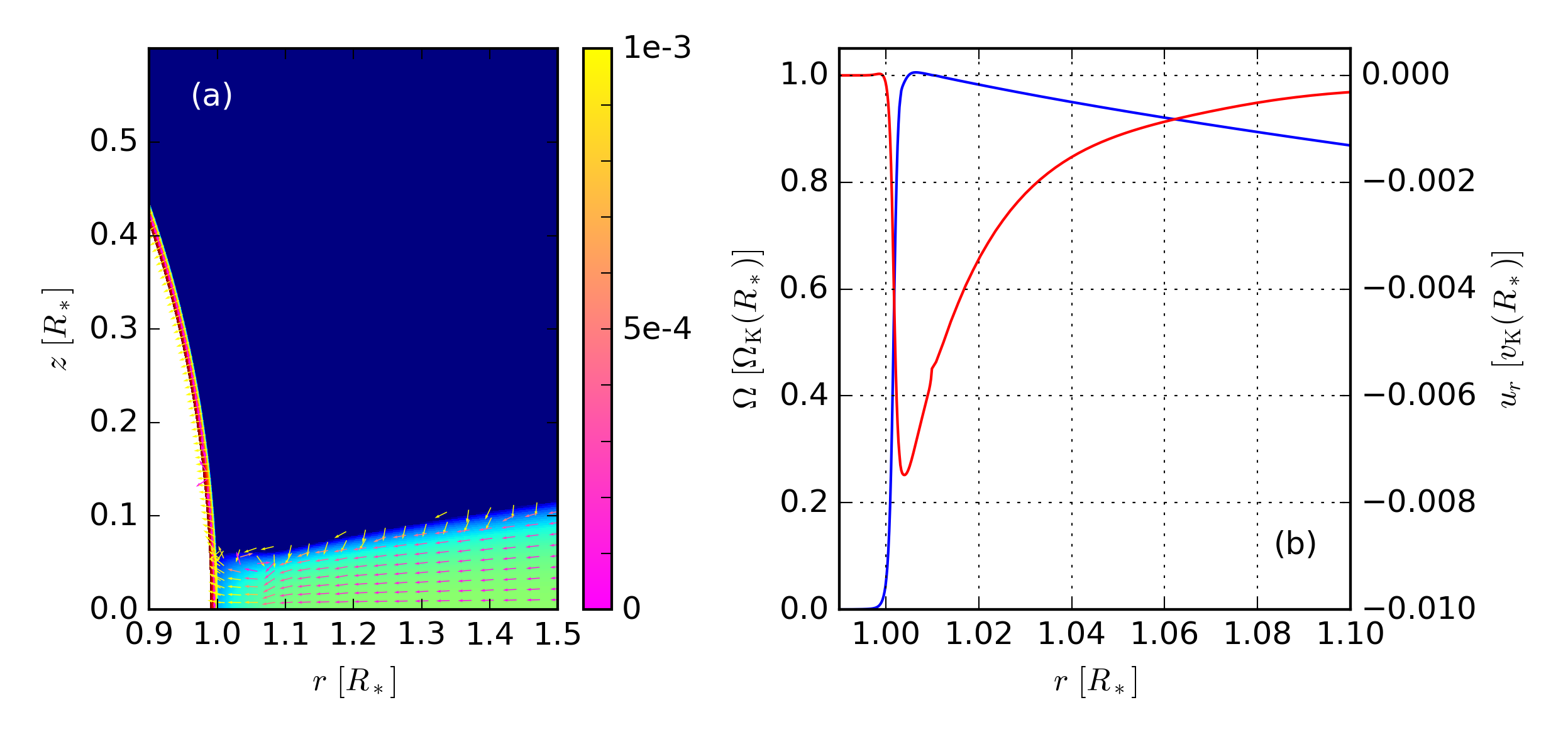}
\caption{\label{fig:v-Omega-ref}%
(a) The density as shown in Fig.~\ref{fig:rho-T-ref} as an orientation for the over plotted velocity vectors. The color-bar indicates the
absolute value of the velocity of each vector in units of $v_\text{K}(R_\ast)$. Only the radial and polar component of the velocity vector have been taken into account.
(b) The angular velocity $\Omega=u_\varphi/r$ (blue) and the radial velocity $u_r$ (red) in the equatorial plane as a function of the
radius $r$. The $y$-axis on the left-hand side refers to $\Omega$ and the $y$-axis on the right-hand side refers to $u_r$. All velocities
are normalized to the Keplerian (angular) velocity at the surface of the star, $v_\text{K}(R_\ast)=\sqrt{GM_\ast/R_\ast}$. Both snapshots
are taken after a time of 688 orbits.
}
\end{center}
\end{figure*}

In Fig.~\ref{fig:v-Omega-ref} (a) the dynamic structure of the star-BL-disk system is displayed. The density is shown as an orientation
for the velocity vectors that visualize the direction and the magnitude of the flow. The disk shows a steady slow inflow towards the
star. The inflow velocity is larger near the disk surface than in the midplane.
As the gas approaches the stellar surface it is considerably sped up. This is clearly visible in Fig.~\ref{fig:v-Omega-ref} (b), where
the radial velocity $u_r$ in the disk midplane is shown in red. The infall of material is accelerated until $r\approx 1.05$ and then
rapidly decelerated as the matter comes to the surface of the star. The peak of $u_r$ corresponds to a Mach number of approximately
$0.25$, so the infall occurs subsonically for our chosen value of $\alpha$, which is important for causality reasons \citep[e.g.,][]{1977MNRAS.178..195P}.
Due to the friction in the disk, the gas loses angular momentum and
its stabilizing force so that the infall velocity increases. As a consequence, the density in this region must decrease accordingly in
order to maintain a constant mass-accretion rate through the disk. The mass flux through concentric spherical shells of radius $r$ is given by
\begin{equation}
        \dot{M} = -4\pi r^2 \rho u_r.\label{eq:mdot}
\end{equation}
Thus, if the velocity becomes larger, the density must decrease. $\dot{M}$ can be derived from Eq.~(\ref{eq:continuity}) assuming a
stationary state ($\partial_t = 0$). Also shown in Fig.~\ref{fig:v-Omega-ref} (b) is the angular velocity $\Omega,$
which describes the rotation of the gas around the star. In the disk, the gas rotates with Keplerian frequency and there is a force balance
between the gravitational force of the star and the centrifugal force of the circular motion. Additional forces  arising through
pressure gradients are, in general, small in the disk and influence $\Omega$ only slightly. As the gas approaches the stellar surface,
it is decelerated and smoothly connects to the rotation rate of the star, which is zero in this case. The deceleration happens very fast
and causes a supersonic velocity drop over a range of less than one percent of the stellar radius. Here, the stabilization of the gas
switches from centrifugal to pressure support. The large gradient of $\Omega$ during the velocity drop is responsible for the heat production
in the BL and drives the high temperature observed in the thermal BL. In contrast, by definition the \emph{dynamical BL} is the region from the stellar
surface to the maximum of $\Omega$ (zero-torque point) and it is extremely small for BLs around WDs. In this case, the width amounts
to $\Delta r = 0.01$. We begin to measure the BL width when the angular velocity deviates by more than $1\%$ from the stellar-rotation rate.
Only the velocities, especially $u_r$ and $u_\varphi$, vary considerably on a scale of the order of the dynamical BL width ($\sim 1\% R_\ast$).
For the other quantities, it is usually not necessary to resolve the dynamical BL in the figures since no rapid changes happen (see e.g., Fig.~\ref{fig:v-Omega-ref} (b)
vs. Fig.~\ref{fig:Omega-T-rot}, lower panels). From an observational point of view, the thermal BL is the essential region. However,
if we compare the dynamics or the width of the BL, we refer to the dynamical BL.

Once the gas has reached the surface of the star, it has lost almost all of its angular and radial momentum and has settled to a dense
shell in the equatorial plane. It is slowly driven to the poles of the star. Above the disk we set the velocities to zero whenever
the density floor is applied. We thus limit the velocity of the inevitably infalling gas since it dissipates energy when it shocks
on the surface of the star or the disk and heats up the low-density regions. Therefore, no velocity arrows are visible in the dark
blue regions of Fig.~\ref{fig:v-Omega-ref}.


\begin{figure}[t]
\begin{center}
\includegraphics[width=.5\textwidth]{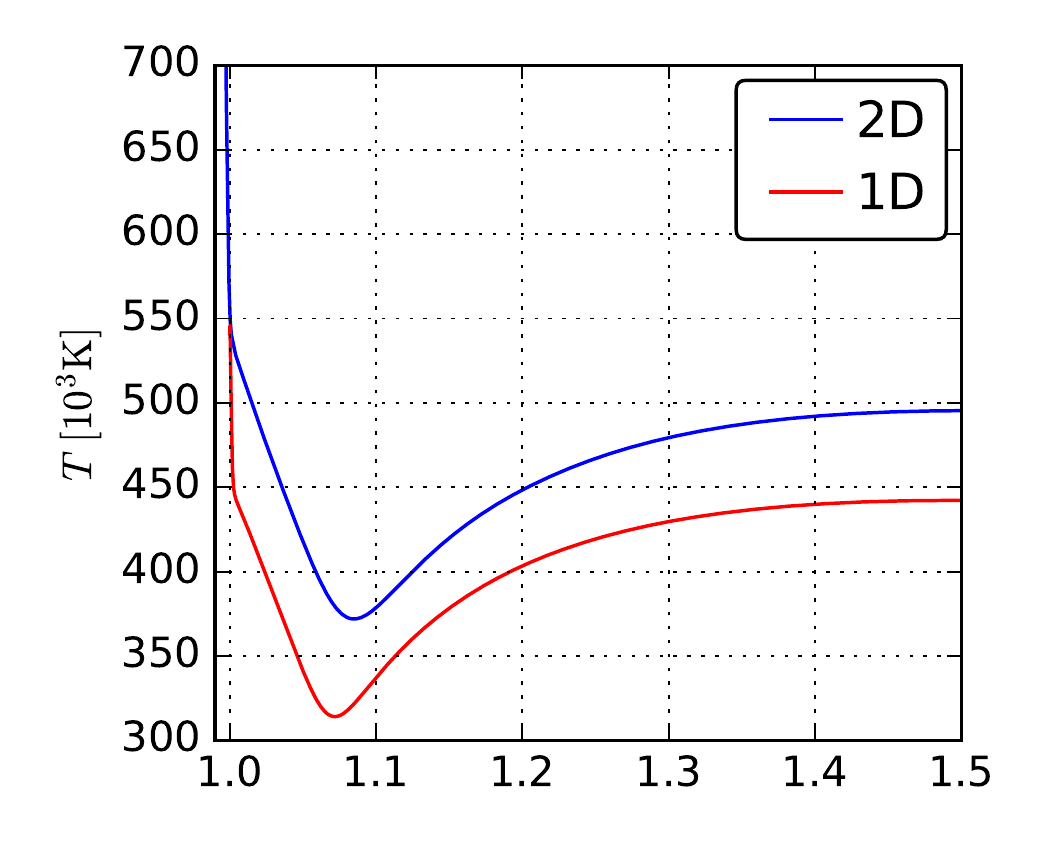}
\caption{\label{fig:1D}%
Midplane temperature $T$ as a function
of radius. We compare the results from 1D (red) and 2D (blue) simulations. A general deviation of
about $10\%$ is due to an insufficient approximation of the vertical optical depth in the 1D approach.}
\end{center}
\end{figure}

We now compare the reference model with the results of the 1D approach where the BL equations in cylindrical coordinates
are vertically integrated and axisymmetry is assumed (see \citealt{2013A&A...560A..56H}). The vertically integrated density (\emph{surface
density}), which is extracted from the 2D data by integrating vertically at constant spherical radius, is similar for the two
approaches. In the disk, the differences are small ($\lesssim 5\%$). The only significant deviation ($\sim 40\%$) is given in the BL where the
1D model predicts a surface density that is  slightly too low. The temperature deviates by approximately $50\,000$ K or $10\%$ throughout the domain, see
Fig.~\ref{fig:1D}. In order to be sure that the reason for this mismatch is not
given by a non fully relaxed 2D midplane temperature, we looked at snapshots of different times and found that in the equatorial
plane, the model is in thermal equilibrium. We conclude that the 1D approximation slightly underestimates the disk temperature.
Apart from the general deviation, the overall trend is well reproduced.
The 1D approach insufficiently approximates the
vertical structure of the disk. This leads to a lower temperature as well as a lower disk height. In the 1D model, thus, the disk
is too thin. This is a consequence of the vertical cooling, which relies on an approximation
of the vertical optical depth analog to a gray atmosphere. If the estimate for the vertical
optical depth is too low, the cooling is more efficient and, hence the temperature in the equatorial plane lower.
The dynamical structure of the 1D model matches the 2D simulation very well.
The shape of $\Omega$ and the width of the BL are virtually identical for the two different approaches.
The infall velocity is
also a close match, though it is slightly larger in the BL since the surface density is smaller (see Eq.~\ref{eq:mdot}).

\begin{figure}[t]
\begin{center}
\includegraphics[width=0.5\textwidth]{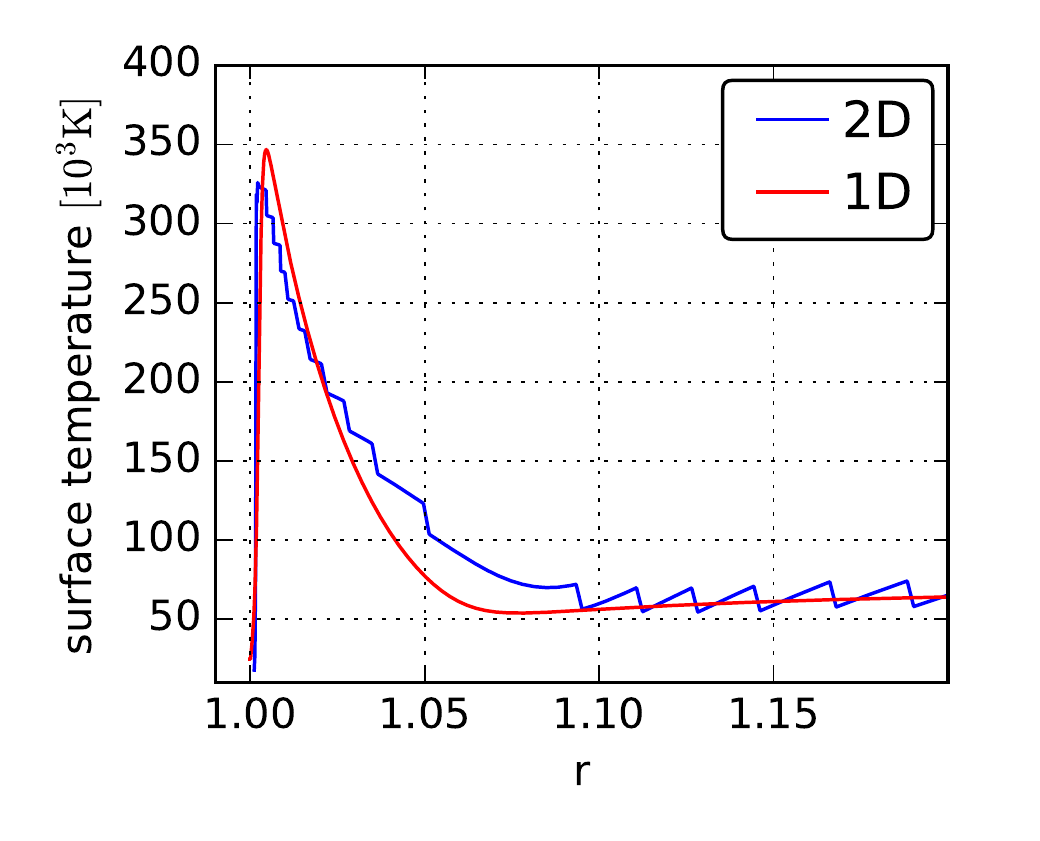}
\caption{\label{fig:Tsurface}%
The surface temperature as a function of radius for the 1D radial model and the 2D model. In the latter, the
radiation temperature $T_\text{rad}$ is calculated following Eq.~(\ref{eq:Trad}) and the surface temperature of the disk and
BL is given by $T_\text{rad}$ at an optical depth of $\tau=1$.
}
\end{center}
\end{figure}

Figure~\ref{fig:Tsurface} shows the surface temperature of the BL and the disk. In the 1D radial approximation,
the surface temperature is derived from the midplane temperature by employing the estimate for the vertical optical depth
$\tau_\text{eff}$ by \citet{1990ApJ...351..632H}:
\begin{equation}
        T_\text{c}^{\text{1D}} = \sqrt[4]{\tau_\text{eff}} T_\text{surface}^{\text{1D}}
.\end{equation}
For the 2D model, the radiation temperature is computed according to
\begin{equation}
        T_\text{rad} = \left(\frac{E}{a}\right)^{1/4},  \label{eq:Trad}
\end{equation}
where $a$ is the radiation constant. The surface temperature is then obtained by evaluating $T_\text{rad}$ in an optical depth
of $\tau=1$. Since the line of constant $\tau=1$ approximately equals the visible surface, the surface temperature is a
measure for the radiation emitted by the system. The 1D and the 2D approaches yield remarkably similar values of $T_\text{surf}$
(Fig.~\ref{fig:Tsurface}). The peak of the 2D surface temperature is slightly shifted towards the star since the BL of
the 2D model lies closer to the star than in the 1D case (see also Fig.~\ref{fig:1D}). The peak is marginally lower and the
descent less steep. In the 2D model, the thermal BL is thus slightly wider and less hot than in the 1D approximation. In the
disk, both models decrease to about $55\,000$ K and the radiation is comparable to the WD. The 2D surface temperature is not
as smooth as the 1D counterpart and shows frequent edges and small spikes. The reason is, that, coming from above, the density
in the disk rapidly rises to the point where $\tau=1$ and the number of grid cells in this region is limited. By increasing
the resolution at the transition to the disk, the smoothness of the profile could be improved.

\subsection{The influence of the WD rotation}

\begin{figure*}[t]
\begin{center}
\includegraphics[width=\textwidth]{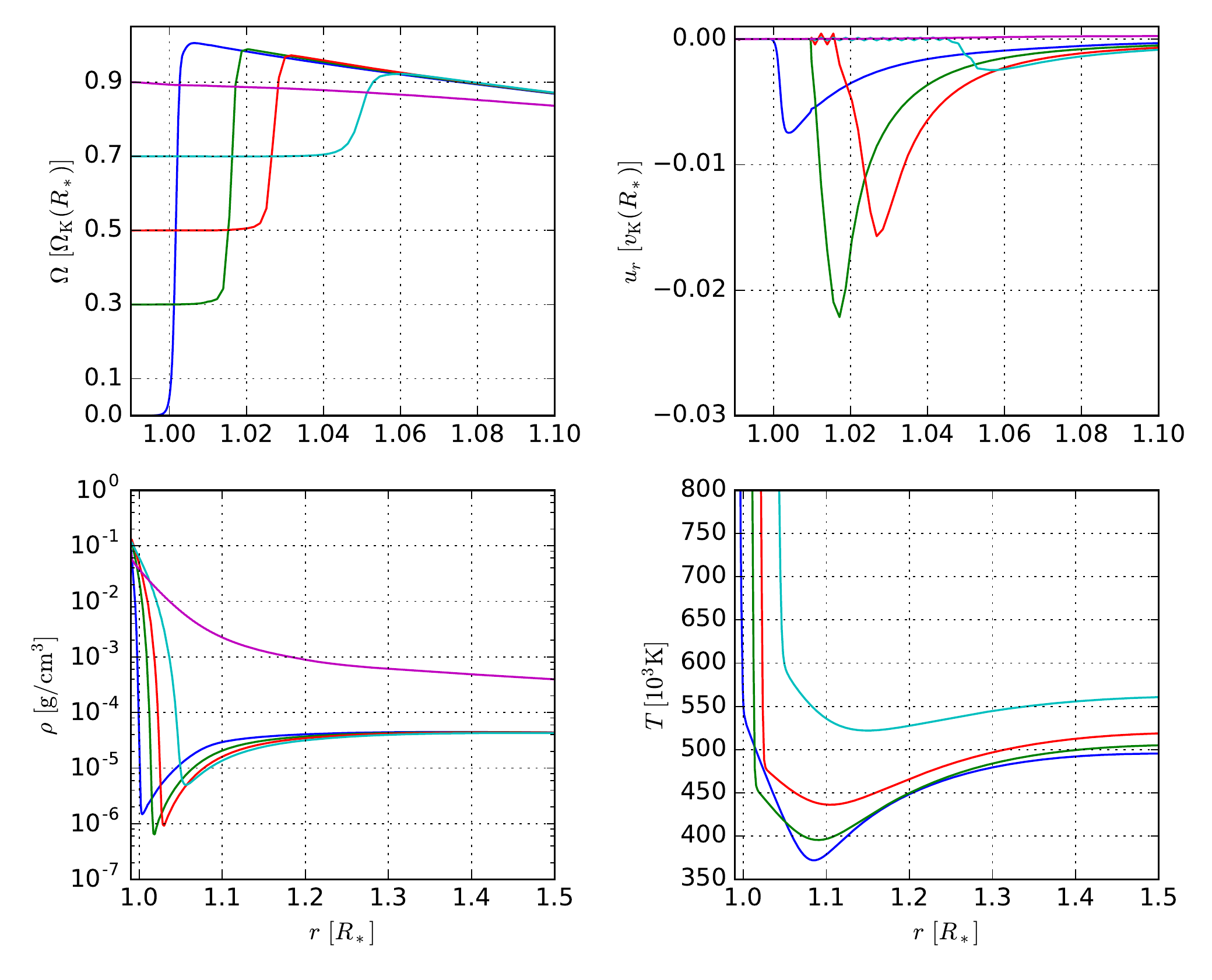}
\caption{\label{fig:Omega-T-rot}%
Angular velocity $\Omega$ and radial velocity $u_r$ (both in dimensionless units) and density $\rho$ and temperature $T$ in cgs units.
All variables show the radial dependency in the equatorial plane. The different colors denote the different stellar rotation rates
of $0, 0.3, 0.5, 0.7$ and $0.9$ times the Keplerian angular velocity at the stellar surface. The snapshots are taken after 688,
251, 248, 263 and 253 orbits. The blue lines correspond to the reference model.
}
\end{center}
\end{figure*}

The rotation rate of the WD is an important parameter since it determines the amount of energy released in the BL. With increasing stellar
rotation rate, the amount of kinetic
energy that is transformed into heat in the BL decreases. The luminosity of the BL decreases quadratically with increasing stellar
rotation rate,
\begin{equation}
        L_\text{BL} = \frac{1}{2}L_\text{acc}\left(1 - \frac{\Omega_\ast}{\Omega_\text{K}(R_\ast)}\right)^2,\label{eq:L-BL}
\end{equation}
where $L_\text{acc}=GM_\ast\dot{M}/R_\ast$ is the total accretion luminosity. Equation~(\ref{eq:L-BL}) can be derived theoretically
\citep{1995ApJ...442..337P} and has been verified numerically \citep{2017arXiv170206726H}.

We considered five different stellar rotation rates ranging from non-rotating up to nearly break-up rotation velocity. The differences
between these models are illustrated in Fig.~\ref{fig:Omega-T-rot} where the midplane radial dependency of the angular velocity, radial
velocity, density and temperature are shown. The angular velocity $\Omega$ decreases smoothly from Keplerian rotation in the disk to
the stellar-rotation rate. The faster the star spins, the more the dynamical BL is shifted to greater radii since the star increases its
equatorial radius due to the centrifugal force imposed by its rotation. For a WD that spins at $70\%$ of $\Omega_\text{K}$, the equatorial
radius increase amounts to already $\sim 4\%$ of the stellar radius. Also, the dynamical BL width becomes larger with increasing stellar
rotation; $\Delta r = 0.01, 0.015, 0.016, 0.025$ for $\Omega_\ast = 0.0, 0.3, 0.5, 0.7 \Omega_\text{K}(R_\ast)$. The reason for the increasing
BL width is the complex interplay of density and temperature, both of which influence the viscosity, which ultimately governs the width
of the BL. In the disk, the angular velocity of all models is nearly Keplerian, with a small deviation due to the outward pointing pressure
gradient, which adds as a stabilizing force.

The infall velocity $u_r$ shows a behavior which is not expected from 1D BL simulations where the maximum of the radial velocity
decreases with rising rotation rate \citep[e.g.,][]{2013A&A...560A..56H}. In contrast, the 2D simulations propose a trend where
the peak of $u_r$ first increases significantly before decreasing again. The turnaround point seems to be in the range of $\Omega_\ast\approx0.3$
and only for a high stellar rotation rate of $\Omega_\ast = 0.7$ is the peak smaller than in the non-rotating model. Here
it would certainly be interesting to run additional models with different $\Omega_\ast$ to investigate the trend in depth. Apart from the
peak behavior, the shape of $u_r$ reflects our expectations. Coming from the disk, $u_r$ is small and increases as the BL is approached. In the
BL, it peaks and then decreases rapidly as the stellar surface is encountered. Although two models peak at a higher infall velocity than the
reference model, the infall is still well sub-sonic.

The unexpected behavior of the rotating models continues to occur in the density and the temperature. While in 1D models there
is a clear trend that the (surface) density increases and the temperature decreases as the star spins up, Fig.~\ref{fig:Omega-T-rot} points
towards a different picture. The density in the BL for the $\Omega_\ast=0.3$ case is smaller than in the non-rotating model. The same applies
for the model where the star rotates at half of the break-up velocity. The trend of each $\rho(r)$ curve, however, is in perfect agreement
with the non-rotating model. The temperature, instead of decreasing with increasing rotation rate, seems to rise at first glance. Important,
though, is the BL temperature which corresponds to the region directly in front of the steep rise as we enter the star. Here, the non-rotating
model has a higher temperature than the $0.3$ and $0.5$ cases, which is in agreement with Eq.~(\ref{eq:L-BL}). The $0.7$ model deviates
from this picture since the equatorial radius increase is already very pronounced. The cause for the increased infall velocity in the rotating
models is unclear. Perhaps the lack of a stabilizing pressure force due to a lower BL temperature is responsible for the accelerated inflow.
As has been mentioned earlier, $\rho$ will adjust to the radial velocity such that the mass flux remains constant, see Eq.~(\ref{eq:mdot}).
This could also explain the lower density in these cases. In general, however, the inflow velocity should be proportional to the loss of
angular momentum and thus be higher for lower $\Omega_\ast$. This point remains to be clarified through further investigation.
The temperature of the $\Omega_\ast=0.9$ model is not visible in Fig.~\ref{fig:Omega-T-rot} since it is larger than $800\,000$ K throughout
the plotted region due to the high density.

A special case is given by the model with a stellar rotation rate of $0.9$. Due to the high rotation rate of the star, the system has reached
a state where matter is fed into the disk by the star and transported outwards by the disk. This phenomenon is called a \emph{decretion disk}.
The radial velocity is pointing outwards throughout the disk. Due to this exotic state of the disk, no BL is formed and the density and
temperature are much higher than in the other cases since the hot stellar material is transported outwards. We find that a decretion disk
is formed in each model that has a rotation rate of $0.9$ and therefore exclude those cases from our further analysis.

For the models with a rotating star we found that the material does not continue its journey to the poles unhindered as was the case for
the non-rotating star. We investigate the polar spreading of the material, which depends on the stellar rotation rate, in detail
in Sec.~\ref{sec:polar-spreading-and-mixing}.


\subsection{Dependency on stellar mass and mass-accretion rate}

\begin{figure*}[t]
\begin{center}
\includegraphics[width=\textwidth]{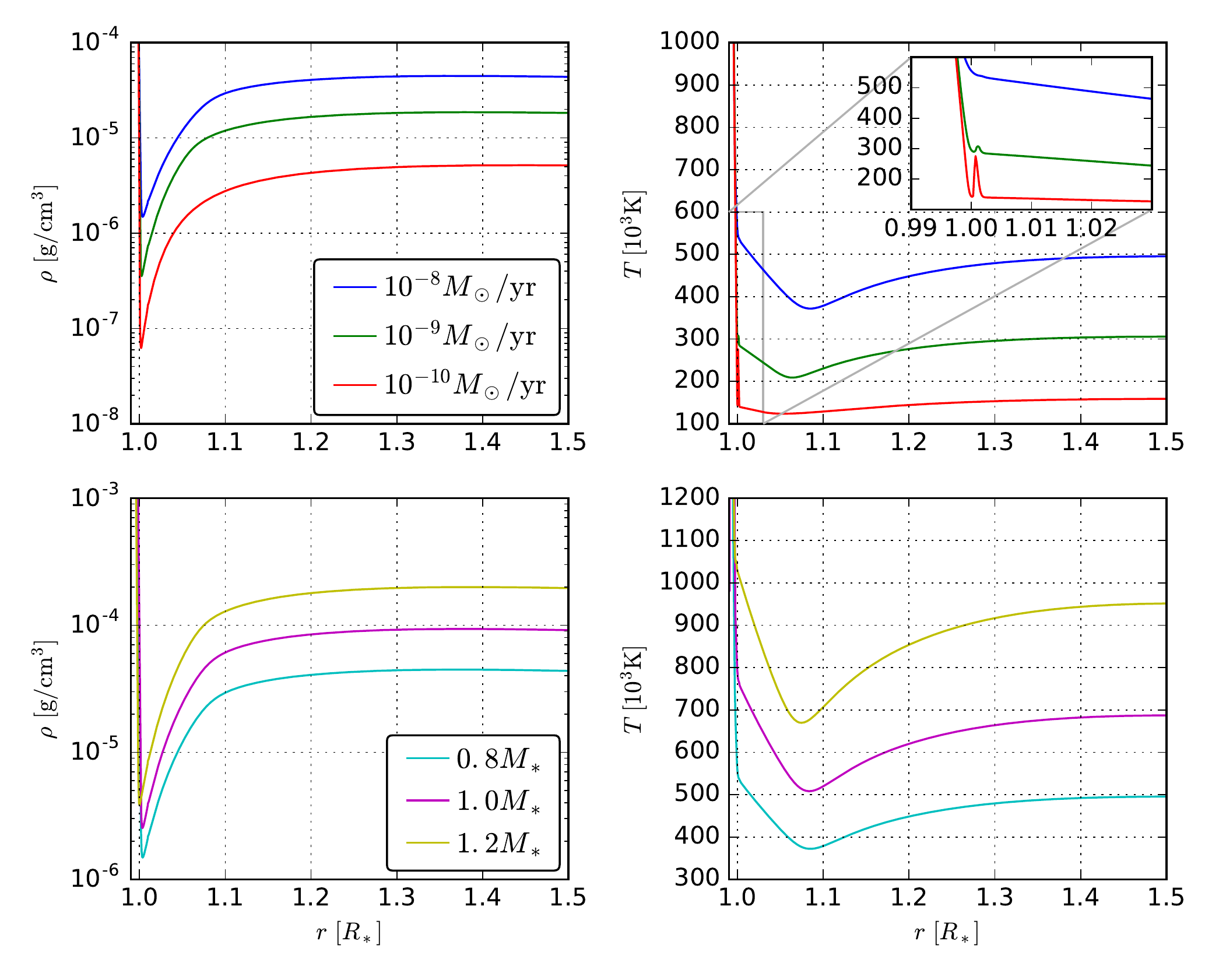}
\caption{\label{fig:Omega-T-mdot}%
Density and temperature in the equatorial plane as a function of radius. In the upper two panels the variables are displayed for three
different mass-accretion rates. The plot of the temperature has an additional zoom box to make the thin peaks in the BL more visible.
The lower two panels show $\rho$ and $T$ for three different stellar masses.
}
\end{center}
\end{figure*}

We now investigate to what extent the stellar mass and the mass-accretion rate of the system influence the structure of the BL. To this
end, we consider the following models (see also Table~\ref{tab:width}): For the $\dot{M}$-study, we leave the stellar mass fixed at $M_\ast=0.8M_\odot$ and compare
three different mass-accretion rates, $10^{-8}, 10^{-9}$ , and $10^{-10} M_\odot/\text{yr}$. We then fix $\dot{M}$ at $10^{-8} M_\odot/\text{yr}$
and change the stellar mass to $0.8,1.0,$ and $1.2 M_\odot$. All models feature a non-rotating WD.

The upper two panels of Fig.~\ref{fig:Omega-T-mdot} show the density $\rho$ and the temperature $T$ in the equatorial plane as a function
of radius for three different mass-accretion rates. With decreasing $\dot{M}$  the density also drops significantly.
A difference of two orders of magnitude in mass-accretion rate leads to a density
drop of one and $1.5$ orders of magnitude in the disk and in the BL, respectively. A similar trend applies to the temperature, which also
decreases with ceasing mass-accretion rate. Here we find a factor of more than three between the temperature in the disk with an accretion
rate of $10^{-10}$ and $10^{-8}$ solar masses per year. While the hottest model reaches over $500\,000$ K in the BL, the coldest model
lies at only  roughly $150\,000$ K, apart from the interesting peak, which we discuss later. The more mass is transported through the
disk, the more massive the disk will grow and hence the density increases with $\dot{M}$. This behavior is also reflected in the disk
height, which increases along with $\dot{M}$. Due to the higher disk density, the depletion of mass in the boundary layer is less severe
for the high $\dot{M}$ case as well. There are two major reasons for  the increasing temperature: On the one hand, the total accretion luminosity
scales linearly with the mass-accretion rate:
\begin{equation}
        L_\text{acc} = \frac{GM_\ast\dot{M}}{R_\ast}    \label{eq:Lacc}
.\end{equation}
It describes the total amount of energy the gas loses in the disk and the BL around a non-rotating star per time. Thus, the higher
the mass-accretion rate, the higher the energy production and also the temperature, since there is simply more material in the
disk, which produces heat through shearing. On the other hand, a higher disk density amplifies the energy dissipation. In the BL, again,
the increase in temperature is more distinct than in the disk.

\begin{table}
\caption{\label{tab:width}%
Width of the boundary layer for a parameter variation of the stellar mass and the mass-accretion rate.
By definition, the BL ranges from the surface of the star to the point where $\partial\Omega(r)/\partial r =0$, that is, where it has a maximum.
}
\begin{center}
\begin{tabular}{l l l l}
        \hline\hline
        $M_\ast$ $[M_\odot]$ & $\dot{M}$ $[M_\odot/\mathrm{yr}]$ & $\Delta r$ $[R_\ast]$ & sim. time $[\text{orbits}]$ \\  \hline
        $0.8$ & $1\times10^{-8}$ & $0.0100$ & 688\\ 
        $0.8$ & $1\times10^{-9}$ & $0.0066$ & 604\\ 
        $0.8$ & $1\times10^{-10}$ & $0.0048$& 395\\ 
        $1.0$ & $1\times10^{-8}$ & $0.0084$ & 395\\ 
        $1.2$ & $1\times10^{-8}$ & $0.0064$ & 387\\ 
        \hline
\end{tabular}
\end{center}
\end{table}

With decreasing mass-accretion rate, an interesting feature arises for the temperature in the BL. In the zoom-in box of Fig.~\ref{fig:Omega-T-mdot},
which magnifies the BL region $[0.99,1.03]$, small peaks for $10^{-9}$ and $10^{-10} M_\odot/\text{yr}$ are visible, which seem to grow
larger with decreasing $\dot{M}$. Those temperature peaks directly reflect the heat production in the BL where the dissipation rate
has its maximum and are also observed in 1D simulations \citep[e.g.,][]{2013A&A...560A..56H}. Since at this location the density is
very low due to the peak of the radial velocity, the matter cannot cool efficiently
enough. The radiation transport depends on the opacity, which strongly depends on the density and the temperature. The lower the mass-accretion rate, the smaller also the density and the transport
of energy away from this region
by radiation, thus cooling this region down,\ becomes more and more difficult. Therefore this peak grows more pronounced with decreasing $\dot{M}$.

In the lower two panels of Fig.~\ref{fig:Omega-T-mdot}, the midplane density and temperature for the three different stellar masses
are presented. The density plot confirms, that a higher stellar mass leads to higher density in the disk and in the BL. The variation
is roughly exponentially and equally pronounced in the disk and the BL. With increasing stellar mass, the radius of the WD decreases
due to the inverse mass-radius relation. This causes a non-linear variation of gravity and the disk height for high-mass WDs is considerably
smaller than for lower-mass WDs. Since the mass-accretion rate remains fixed, the amount of mass transported through the disk also
remains constant and thus the density increases with $M_\ast$. According to Eq.~(\ref{eq:Lacc}) and in combination with the higher
density, the temperature also increases strongly. For the case of a very massive $1.2$ solar mass WD, the one million K temperature
mark is reached. Again, the temperature in the BL grows stronger than in the disk with increasing $M_\ast$. A peak in the BL is not
visible due to the high mass-accretion rate and the effective radiative cooling. For reference, we state the radii for the WDs of
masses $0.8,1.0,$ and $1.2 M_\ast$ which amount to $7.1\times10^8, 5.6\times10^8$ , and $4\times10^8$ cm, respectively.

The dynamical structure does not significantly change with the stellar mass or the mass-accretion rate and can be described by the
general flow structure of the reference model (see Sec.~\ref{sec:basic-structure}). The width of the BL, however, depends on $\dot{M}$ and
$M_\ast$ as can be extracted from Table~\ref{tab:width}. There is a clear trend for all stellar masses that the BL shrinks with
decreasing mass-accretion rate. The width further decreases with increasing stellar mass. High mass-accretion rates lead to broader BLs
since the density and the temperature are larger and consequently the BL hotter. This effect is reinforced by the inefficient radiation
transport in the BL for low mass-accretion rates, which prevents a more distinct expansion. When considering the stellar masses, however,
higher densities and temperatures cause quite the opposite, namely a shrinking BL width. Here, we have to take into account the increased
gravitational pull, which dominates in this case and drags the disk as close as possible to the stellar surface. Several factors contribute
to the width of the BL and it is therefore difficult, to come to a unique conclusion without running simulations.
The radial velocities and Mach numbers are almost identical, apart from the slight radial shift due to the different BL widths. We also
observe an unhindered, slow polar spreading of the disk material shell in all models.



\subsection{Polar spreading and mixing}\label{sec:polar-spreading-and-mixing}

\begin{figure*}[t]
\begin{center}
\includegraphics[width=\textwidth]{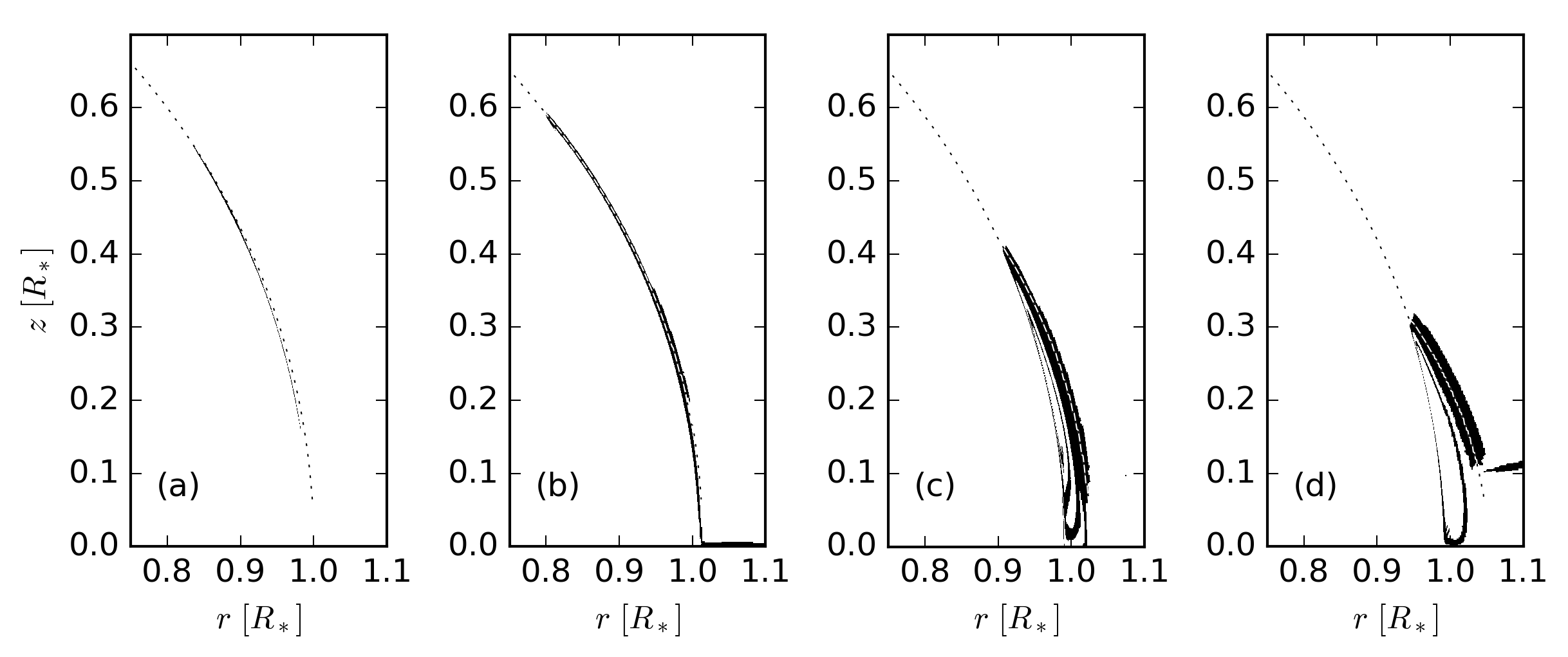}
\caption{\label{fig:spreading1}%
Black and white image of the tracer, which has been inserted in the equatorial plane in order to visualize the mixing and polar spreading
of the disk material of a $0.8 M_\odot$ WD with $\dot{M}=10^{-8}M_\odot/\text{yr}$. The plots differ in the stellar-rotation rate,
which is given by $0.0$ (a, orbit 688), $0.3$ (b, orbit 251), $0.5$ (c, orbit 248) and
$0.7 \Omega_\text{K}(R_\ast)$ (d, orbit 263). The dotted black line denotes the optical depth of $\tau=2/3$.
}
\end{center}
\end{figure*}

In Fig.~\ref{fig:spreading1},
the destination of the disk material on the stellar surface is visualized. To this purpose, we have added a tracer in the
equatorial plane just short of the stellar surface in the simulations. The tracer behaves like a scalar quantity for which an
advection equation is solved. It thus illustrates in which direction the material from the disk is going. Along with the tracer
we have added the line of constant $\tau=2/3$ with a dotted black line that is approximately the visible surface of the star.
Four different stellar rotation rates are given.

\begin{figure}[t]
\begin{center}
\includegraphics[width=0.5\textwidth]{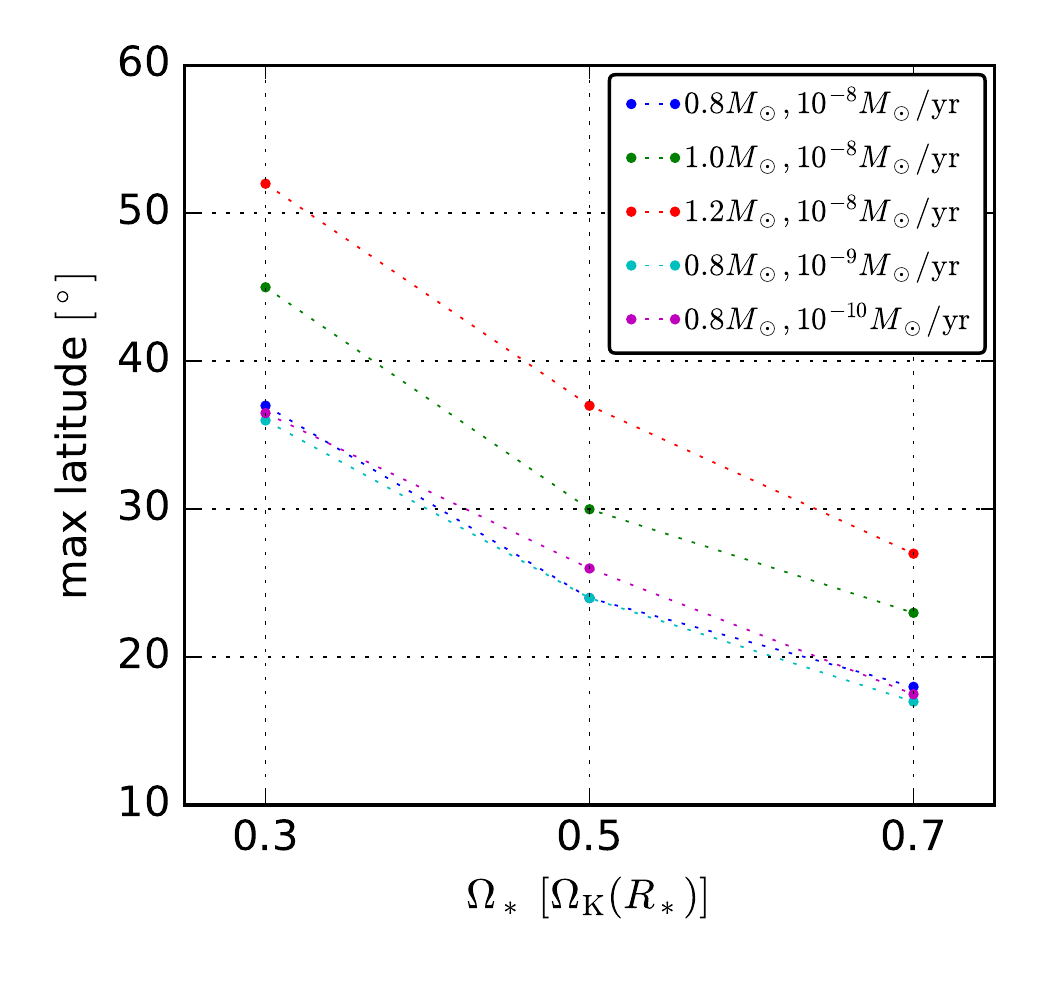}
\caption{\label{fig:spreading3}%
The maximum latitude reached by the disk material on its way to the pole as a function of the stellar-rotation rate. The
different curves correspond to different choices of stellar mass and mass-accretion rate (see legend). The maximum latitude
angle is measured towards the equatorial plane and is given in degrees.
}
\end{center}
\end{figure}

In Figure~\ref{fig:spreading1} (a) the situation is shown for the reference model with the non-rotating WD. As was mentioned
earlier, the disk material is gathered in a very thin shell that slowly moves to the poles. 
Since the WD is not rotating and the material has been completely depleted of its angular momentum near the midplane, the
surface of the WD is an equipotential area. The gas is pushed up towards the poles by the pressure that has developed in the
midplane at $r\approx 1$ due to the deceleration of the disk. On the WD surface, the movement is slowed down by friction. If the
viscosity $\nu_\text{const}$ of the WD is reduced, we observe that the material reaches the poles faster than in the reference
model. In these cases, we can actually follow the simulation for the time that is needed for the gas to reach the pole. However,
the choice of an overly small $\nu_\text{const}$ is problematic since it prevents the BL from building up in the first place. For the rotating models it is especially important that $\nu_\text{const}$ is large enough so that the WD rotates like a solid body. We
have found that $\nu_\text{const}=10^{12}\text{cm}^2/\text{s}$ is a good choice for this purpose.

The picture changes when the star begins to rotate. In this case, there is a centrifugal barrier that initially prevents the
gas from leaving the midplane. However, the pressure force in the equatorial plane is large enough to overcome this barrier
and push the material towards the pole. At a certain latitude, an equilibrium of pressure and centrifugal force is established
and the movement is halted. This situation is illustrated in panels (b) - (d) of Figure~\ref{fig:spreading1}, where the WD rotates with $30\%, 50\%,$ and
$70\%$ of $\Omega_\text{K}(R_\ast)$. Since with increasing stellar rotation rate the centrifugal force increases and the
pressure force decreases, the maximum latitude shrinks considerably.

The mixing of the stellar and disk material also becomes more pronounced with increasing rotation rate. In all three cases
where the WD spins at a non-zero rate, the $\tau = 2/3$ is engulfed in a mixture of material from the disk and the star. Therefore
it is to be expected that the radiation of the star is reprocessed in regions partly dominated by disk material; material that might have a
composition different from the stellar mixture in some cases. This situation occurs, however, only for a part of the star and two caps at the poles, which
increase in size as the star spins up and remain untouched by the disk material. 

Figure~\ref{fig:spreading3}, where the maximum latitude as a function of $\Omega_\ast$ is shown, visualizes how the spreading is 
influenced by $M_\ast$ and $\dot{M}$. The trend of the polar spreading of material decreasing with stellar
radius from Fig.~\ref{fig:spreading1} is confirmed by Fig.~\ref{fig:spreading3} for various combinations of stellar mass and mass-accretion rate. The gas faces a centrifugal barrier when traveling to the poles due to the fact that it retains an increasing amount
of angular momentum with growing stellar rotation rate when reaching the stellar surface. The viscosity on the stellar surface on
regions off the equatorial plane is far too low to decelerate the material efficiently and rid it of its excess angular momentum.
This also explains the unhindered slow movement of the disk gas in the non-rotating case.

Figure~\ref{fig:spreading3} further suggests that an increase in stellar mass leads to a considerable growth of the maximum latitude.
The rotational velocity on the surface of the star is proportional to $\sqrt{M_\ast/R_\ast}$ and thus is growing with increasing
stellar mass. This means that a point on the equator of a $1.2 M_\odot$ WD that rotates with $30\%$ of the break-up velocity moves
at almost $2000 \text{km}/\text{s}$, whereas for a $0.8 M_\odot$ WD the velocity is only about $1200 \text{km}/\text{s}$. The important
quantity for the polar spreading, however, is the angular momentum, which scales $\propto \sqrt{M_\ast R_\ast}$ on the surface and
thus decreases with increasing stellar mass due to the inverse mass-radius relation of WDs. The material can therefore reach higher
latitudes on the WD. The mass-accretion rate, on the contrary, does not seem to influence the polar spreading at all. The small deviations
between the blue, cyan, and magenta curves in Fig.~\ref{fig:spreading3} are probably due to inaccuracies in measurement. There is
therefore no increased pressure force to drive the material further to the poles for increasing mass-accretion rate. The angular
momentum of the gas remains the key issue in the polar spreading. It would be interesting to investigate the shape of the $\Omega_\ast$
dependence of the maximum latitude further, however, with only three data points per curve we feel that an interpretation beyond
the general trend is too speculative. We leave that issue for future studies.

In all of the simulations we have performed, the disk material is decelerated in the midplane and the radiation of the BL
emerges in proximity to the stellar equator. The situation depicted in Fig.~\ref{fig:Erad-ref}  also holds true for increasing
stellar rotation rate $\Omega_\ast$ and only the amount of dissipated energy decreases. The disk material that finally
spreads towards the poles has the same rotational velocity as the stellar surface. Therefore, we do not observe a spreading
layer (SL) as mentioned in \citet{1999AstL...25..269I,2010AstL...36..848I}, where the rotating material first spreads on the
star and is decelerated and radiates within two rings above and beyond the equator. The latitude of these rings depends on the
mass-accretion rate. However, we also do not detect any such changes in the BL when the mass-accretion rate is varied
(for instance Fig.~\ref{fig:spreading3}).

\section{Summary and conclusion}\label{sec:conclusion}

For this publication we have performed 2D simulations of the non-magnetic BL around WDs in CVs in a spherical geometry
assuming axisymmetry. The hydrodynamical model based on the Navier-Stokes equations is extended by an additional equation for
the radiation energy, which is closed using the flux-limited diffusion approach. We employ Kramer's opacity, which suits the temperatures
found in the inner disk around WDs and utilize a modified $\alpha$-prescription for the viscosity, which also takes into
account the radial scale height in the BL. A total of 45 models have been
prepared in a complex way (see Sec.~\ref{sec:solution-strategy}) and run with varied $M_\ast$ ($0.8, 1.0, 1.2 M_\odot$),
$\dot{M}$ ($10^{-8}, 10^{-9}, 10^{-10} M_\odot/\text{yr}$) and
$\Omega_\ast$ ($0.0, 0.3, 0.5, 0.7, 0.9 \Omega_\text{K}(R_\ast)$). Each model features a high resolution ($\sim 1.1$ million cells)
and has been run for several hundred orbits, both of which are unprecedented, especially when considering the elaborate physics
included in the model. The downside of this advanced setup are simulation times of $\gtrsim 1$ week with 560 parallel cores on the
most recent cluster hardware. However, this can be coped with thanks to the availability of high-performance computing centers in
Baden-Württemberg, Germany.

We found that in the BL the angular velocity of the gas decreases smoothly from the Keplerian rotation in the disk towards the
non-rotating stellar surface. Due to the loss of stabilization by angular momentum, the infall velocity of the
gas increases and peaks in the BL. This goes along with a severe depletion of mass which is characteristic for the BL and suggests
the picture of a bottleneck. Since the gas loses a great amount of energy before coming to rest on the star, very high temperatures
of $\sim 550\,000$ K for a $0.8 M_\ast$ WD with a mass accretion of $10^{-8} M_\ast/\text{yr}$ are reached. Whereas the
dynamical BL, that is, the region where $\Omega$ drops to $\Omega_\ast$, is very small ($\lesssim 1\% R_\ast$), the area over which
the dissipated energy is radiated away is considerably larger ($\sim 10\% R_\ast$). This is of particular interest for the observational
appearance of the BL, since a wider region means less hard radiation. The radiative flux further reveals that the hottest
(i.e., closest to the star) part of the disk also contributes to the thermal BL. This might lead to an overestimated $\Omega_\ast$
since the BL dissipation is assumed slightly too large when comparing observations with synthetic BL models.

By increasing the stellar-rotation rate, the general structure of the BL does not change. The angular velocity still smoothly
connects to the surface velocity and the density and temperature in the BL are small and large, respectively. The dynamical width,
however, increases due to the changing viscosity which is governed by the temperature. One open question remains with
the infall velocity that, against our expectations, increases with rising $\Omega_\ast$ at first and decreases only for
$\Omega_\ast \gtrsim 0.3$. An interesting case is found for the high stellar-rotation rate of $0.9\Omega_\text{K}(R_\ast)$ throughout
the $M_\ast, \dot{M}$ parameter space. Here, a decretion disk is formed and material from the star is transported outwards
through the disk. In general, the stellar-rotation rate is an important parameter since it determines the amount of energy
liberated and radiated away in the BL and can in principle be identified by analyzing the BL luminosity from observations. There is,
however, an ambiguity in that different choices of $\Omega_\ast, M_\ast$ , and $\dot{M}$ can lead to similar luminosities, which
complicates this process. One way to circumvent this difficulty entails measuring the BL X-ray luminosity along with the bolometric
luminosity since \citet{2017arXiv170206726H} has recently shown that the X-ray luminosity scales $\propto \Omega_\ast^3$ as opposed to the
$\propto \Omega_\ast^2$ dependency of the total luminosity.

The mass-accretion rate and the stellar mass both affect the BL such that a higher value leads to an increase in density and
temperature. In the first case, the higher amount of mass in the disk due to the higher accretion rate and hence the increased
dissipation is responsible for the rise in density and temperature. In the other case, a higher $M_\ast$ intensifies the gravitational
pull on the material. A process which is supported by the inverse mass-radius relation of WDs. The width of the BL decreases
with increasing stellar mass or decreasing mass-accretion rate for the same reasons. The height of the disk also follows this trend.
Intriguingly, with decreasing mass-accretion rate a thin but pronounced temperature peak within the dynamical BL becomes visible
due to an inefficient radiative cooling under these conditions. At these mass-accretion rates, the BL starts to become
optically thin. The value matches other studies, which give a threshold of about $10^{-10}M_\odot/\text{yr}$ for an optically thin
BL \citep[e.g.,][]{1987MNRAS.227...23W}.  Whether or not the peak becomes higher and more
pronounced with further decreasing mass-accretion rate, as well as whether or not  this finally results in soft and hard X-ray emission
from the BL, should be investigated through
future studies.

Due to its complexity and resource consumption, the BL is still frequently treated in a 1D approximation
\citep[see for instance][]{2013A&A...560A..56H,2017arXiv170206726H}. We found that this approach is well justified if the right
conclusions are drawn from the results. Insight into the vertical structure or the interplay between BL and star is clearly
inaccessible by this method. The radiation and luminosity, however, can be deduced even from 1D models, especially
since the surface temperatures are almost identical. One must bear in
mind, though, that the 1D model predicts an overly low midplane temperature and disk height due to an insufficient approximation of the
vertical optical depth. Apart from this drawback and a small artificial radial shift, the midplane profiles
are amazingly similar between the two different approaches. Because of the vastly reduced simulation time ($\sim$ hours), 1D
models are to be preferred for luminosity calculations, for instance.

The results further indicate that the polar spreading, that is, the
motion of the disk material in the direction of the poles of the star, depends on the rotation rate. For a non-rotating star,
a dense shell creeps towards the poles and does not come to rest before reaching them.
With increasing rotation rate, the maximum latitude reached by the gas recedes due to the angular momentum that the
gas is not able to get rid of on the slippery stellar surface at higher latitudes. Also, the mixing with the stellar material
becomes more pronounced with increasing stellar rotation and the gas from the disk comes to rest on and around the observable
surface. As a consequence, the stellar radiation might be reprocessed in a mixture of star and disk material and carry away
the spectral features of both. We note, however, that in our setup, differences in the mean molecular weight
of the accreted and WD material and gravitational settling are not taken into account.
While variations in mass-accretion rate have little or no influence on the polar spreading,
a higher stellar mass leads to higher maximum latitudes. The angular momentum at the stellar surface for a fixed rotation
rate (e.g., $0.3 \Omega_\text{K}(R_\ast)$) decreases with increasing stellar mass due to the inverse mass-radius relations
of WDs. Thus for regular relations, as for instance in young stars, the maximum latitude is likely to decrease with increasing
stellar mass. Our simulations show no indication of a deceleration of the disk material above (or beyond) the stellar
equator for any chosen parameter set. Therefore, we can rule out the concept of the SL for our setup. It remains to be
investigated whether or not this picture changes when a different mechanism for the viscosity is employed \citep[e.g.,][]{2012ApJ...752..115B}.

%
In general, our results are in good agreement with \citet{1991A&A...247...95K} who performed simulations of the BL around
a solar mass WD. We find a slightly smaller BL width for comparable parameters. This is, however, probably a consequence
of small differences in the applied viscosity prescription. Furthermore, we do not detect the extremely hot corona ($\sim10^8$ K)
above the disk found in \citet{1991A&A...247...95K}.
In contrast, this region has a temperature of only $\sim 10^4$ K, which is probably too low and a consequence
of the simplified radiation treatment in the optically thin corona.
The author also mentions that if a turbulent ansatz for the viscosity is
taken (similar to Eq.~\ref{eq:viscosity-nu}), the flow shows a strong time variability with an irregular and eddy-like
structure. The question is raised of whether this is a consequence of inconsistencies with initial conditions or a general feature
of accretion disk flows. Analyzing our simulations, we believe now that this is a transient phenomenon due to some involuntary
initial perturbations, which disappears after several orbits of simulation time. A higher numerical resolution also mitigates
this problem. We can, however, not rule out that, for a specific choice of parameters, a real physically unstable state develops.
Instabilities are more likely found in the disk plane though \citep[e.g.,][]{2015A&A...579A..54H}.
More recently, \citet{2009ApJ...702.1536B} conducted simulations of the BL around a WD with a strong focus on the outer
stellar layers. They varied the $\alpha$ parameter and found optically thick BLs that extend to more than $30^\circ$ to
either side of the disk plane after a short time. While we also find optically thick BLs for $\alpha=0.01$ and high
mass-accretion rates, the inflation of the BL is absent in our simulations and both the disk and the BL are rather thin,
as is expected around a compact object. The unphysical treatment of the dissipated energy most likely causes the
behavior observed in \citet{2009ApJ...702.1536B}. We also did not detect gravity waves or Kelvin-Helmholtz instabilities
although the outer stellar layers are included and are highly resolved in our setup.


The non-magnetic BL has been investigated in a 1D radial approximation \citep{2013A&A...560A..56H}, in a 2D cylindrical
approach in order to examine the instabilities in the disk plane \citep{2015A&A...579A..54H}, and finally in the current paper in a 2D spherical
geometry that reveals the vertical structure. The mammoth task that still remains to be accomplished is the combination of
the latter two. Full 3D simulations will reveal how the AM transport by the BL instability influences the
vertical structure and vice versa. Sadly, this lies beyond what is computationally feasible at this time since a high resolution
in all three coordinate directions is necessary (especially in the azimuthal direction) and particularly the 3D
radiation transport will slow down the simulations considerably. There are, however, other open questions in connection
with the BL such as the influence of magnetic fields or the role of the BL around protoplanets in disks which can be tackled
in a 1D or 2D approach.

\begin{acknowledgements}
Marius Hertfelder received financial support from the German National Academic Foundation (Studienstiftung des deutschen Volkes).
This work was performed on the computational resource ForHLR II funded by the Ministry of Science, Research and the Arts
Baden-Württemberg and DFG ("Deutsche Forschungsgemeinschaft").
Preprocessing of the production runs has been carried out on the BwForCluster BinAC and the authors acknowledge support by the
state of Baden-Württemberg through bwHPC and the German Research Foundation (DFG) through grant no INST 39/963-1 FUGG.
MH wants to thank Hartmut Häfner of the Steinbuch Centre for Computing (SCC) at KIT and Volker Lutz at the Zentrum für Datenverarbeitung (ZDV) in Tübingen,
for their support.
We further made use of \texttt{matplotlib} \citep{Hunter:2007}, a python module for data visualization.
We also thank the referee for his/her constructive comments which helped to improve this paper.
\end{acknowledgements}

\bibliographystyle{aa}
\bibliography{myrefs}
\end{document}